\documentclass[aps,prl,showpacs,reprint,superscriptaddress, nofootinbib]{revtex4-1}
\usepackage{graphicx}
\usepackage{amsmath}
\usepackage{amssymb}
\usepackage{dsfont}
\usepackage{braket}
\usepackage[normalem]{ulem}
\usepackage{wasysym}
\usepackage{placeins}
\usepackage{comment}

\usepackage{float}

\usepackage{esvect}

\usepackage{dcolumn}
\usepackage{bm}

\usepackage{color} 

\newcommand{\delete}[1]{}
\newcommand{\replace}[2]{#2}
\newcommand{\add}[1]{\textcolor{black}{#1}}

\newcommand{\wj}[1]{\textcolor{blue}{WJ: #1}}
\usepackage{hyperref}
\hypersetup{
colorlinks=true,final=true,
        linkcolor=blue,
        citecolor=blue,
        filecolor=blue,
        urlcolor=blue,
}
\begin{document}

\title{Frequency-resolved decoherence spectroscopy of a semiconductor charge qubit coupled to a high-impedance resonator}


\author{E.~Al-Tavil}
\email{ealtavil@phys.ethz.ch}
\affiliation{Department of Physics, ETH Zurich, CH-8093 Zurich, Switzerland}
\affiliation{Quantum Center, ETH Zürich, CH-8093 Zürich, Switzerland}

\author{W.~Jang}
\affiliation{Center for Quantum Science and Engineering, Institute of Physics, École Polytechnique Fédérale de Lausanne (EPFL), 1015 Lausanne, Switzerland}
\affiliation{Hybrid Quantum Circuits Laboratory (HQC), Institute of Physics, École Polytechnique Fédérale de Lausanne (EPFL), 1015 Lausanne, Switzerland}

\author{D.~J.~van~Woerkom}
\altaffiliation{Current address: Microsoft Quantum Lab Delft, Lorentzweg 1, 2628 CJ Delft, The Netherlands}
\affiliation{Department of Physics, ETH Zurich, CH-8093 Zurich, Switzerland}

\author{V.~Maisi}
\affiliation{NanoLund and Solid State Physics, Lund University, Box 118, 22100 Lund, Sweden}

\author{S.~Bosco}
\affiliation{QuTech and Kavli Institute of Nanoscience, Delft University of Technology, Lorentzweg 1, 2628 CJ Delft, The Netherlands}

\author{J.~A.~Krzywda}
\affiliation{Lorentz Institute and Leiden Institute of Advanced Computer Science, Leiden University, P.O. Box 9506, 2300 RA Leiden, The Netherlands}

\author{J.~Danon}
\affiliation{Department of Physics, Norwegian University of Science and Technology, NO-7491, Trondheim, Norway}

\author{C.~Reichl}
\affiliation{Department of Physics, ETH Zurich, CH-8093 Zurich, Switzerland}

\author{W.~Wegscheider}
\affiliation{Department of Physics, ETH Zurich, CH-8093 Zurich, Switzerland}

\author{T.~Ihn}
\affiliation{Department of Physics, ETH Zurich, CH-8093 Zurich, Switzerland}
\affiliation{Quantum Center, ETH Zürich, CH-8093 Zürich, Switzerland}

\author{K.~Ensslin}
\affiliation{Department of Physics, ETH Zurich, CH-8093 Zurich, Switzerland}
\affiliation{Quantum Center, ETH Zürich, CH-8093 Zürich, Switzerland}

\author{A.~Wallraff}
\affiliation{Department of Physics, ETH Zurich, CH-8093 Zurich, Switzerland}
\affiliation{Quantum Center, ETH Zürich, CH-8093 Zürich, Switzerland}

\author{P.~Scarlino}
\email{pasquale.scarlino@epfl.ch}
\affiliation{Center for Quantum Science and Engineering, Institute of Physics, École Polytechnique Fédérale de Lausanne (EPFL), 1015 Lausanne, Switzerland}
\affiliation{Hybrid Quantum Circuits Laboratory (HQC), Institute of Physics, École Polytechnique Fédérale de Lausanne (EPFL), 1015 Lausanne, Switzerland}

\date{\today}


\begin{abstract}
Superconducting resonators coupled to semiconductor quantum dots provide a powerful platform to investigate light–matter interaction and decoherence mechanisms in solid-state quantum systems. Here we study a hybrid circuit quantum electrodynamics architecture consisting of a GaAs double-quantum-dot charge qubit capacitively coupled to a high-impedance, frequency-tunable SQUID-array resonator. By tuning the qubit transition frequency over the range $\omega_\mathrm{q}/2\pi \sim 3$–$6$~GHz, we perform frequency-resolved decoherence spectroscopy of the charge qubit across a broad energy window. 
Time-resolved measurements enable us to disentangle relaxation and pure dephasing processes and to identify distinct decoherence regimes as a function of qubit frequency. We find that at lower frequencies ($\leq 4.5$~GHz) dephasing dominates the qubit linewidth, whereas at higher frequencies energy relaxation becomes the leading contribution. The measured frequency dependence of the relaxation rate exhibits a cubic scaling, 
consistent with charge-qubit decay dominated by coupling to a piezoelectric phonon bath and providing frequency-resolved access to the corresponding phonon-induced spectral density.
Our results show that hybrid semiconductor--superconducting circuits can serve as sensitive spectroscopic tools to probe microscopic decoherence mechanisms relevant for a wide range of hybrid quantum devices.
\end{abstract}

\maketitle

Superconducting microwave resonators constitute a central building block of circuit quantum electrodynamics (cQED) \cite{Blais2021}, enabling coherent control, high-fidelity readout, and photon-mediated coupling between quantum systems \cite{Wallraff2004,Majer2007}. Extending this architecture to semiconductor quantum dots has created hybrid platforms in which microwave photons interact with gate-defined electronic states~\cite{burkard2020,Viennot2014,petersson_circuit_2012,Frey2012}. In double quantum dots (DQDs), charge qubits couple strongly to resonator electric fields because of their large electric dipole moment, enabling strong charge--photon coupling, dispersive charge readout, and access to regimes of ultra strong light--matter interaction~\cite{Mi2017,Stockklauser2017,Bruhat2018,Scarlino2019,scarlino2022,dePalma2024,janik2025,oppliger2026}. The same electric dipole moment, however, makes charge qubits highly sensitive to fluctuations of the electrostatic environment, resulting in limited coherence times~\cite{Petersson2010,hayashi2003,friesen2017,hu2006,Thorgrimsson2017,kim2015}.

Spin qubits provide a complementary route, offering substantially longer intrinsic coherence times and compatibility with scalable semiconductor fabrication~\cite{Hanson2007,Burkard2023}. Their direct magnetic-dipole coupling to microwave photons is weak~\cite{RashbaSheka1991}, and coherent spin--photon interfaces therefore rely on spin--charge hybridization, engineered for example by magnetic-field gradients~\cite{mi2018b,samkharadzeStrong2018}, exchange interactions~\cite{landig2018}, or spin--orbit coupling~\cite{ungerer2024,Yu2023}. This strategy has enabled microwave-cavity detection of spin blockade~\cite{Landig2019}, fast dispersive spin readout~\cite{noirot2026,dijkema2025}, and photon-mediated spin--spin interactions~\cite{borjans2020,dijkema2025,Harvey-Collard2022}. At the same time, the charge admixture required for coupling makes spin--photon devices susceptible to charge-noise-induced decoherence~\cite{dijkema2025,croot2020,noirot2026}. Charge qubits therefore provide a sensitive and well-controlled probe of the microscopic noise and dissipation channels that also affect more complex hybrid spin--charge architectures~\cite{Petersson2010,hayashi2003,shi2013,Thorgrimsson2017,noirot2026,kim2015}.

A key ingredient in these experiments is the use of superconducting resonators with characteristic impedance well above the conventional $50\,\Omega$ environment~\cite{Masluk2012,Samkharadze2016}. High-impedance resonators, realized for example using SQUID arrays~\cite{Castellanos2007,Altimiras2013,Stockklauser2017} or high-kinetic-inductance superconductors~\cite{Annunziata2010,Samkharadze2016,Niepce2019}, increase the vacuum electric-field fluctuations and thereby enhance the coupling to semiconductor charge and spin states~\cite{Stockklauser2017,Scarlino2019,samkharadzeStrong2018,landig2018}. When combined with frequency tunability, such resonators also provide a way to probe decoherence mechanisms over a broad energy range while maintaining sufficient dispersive readout signal.

In this work, we use a frequency-tunable high-impedance SQUID-array resonator capacitively coupled to a GaAs DQD charge qubit to perform frequency-resolved decoherence spectroscopy over the qubit frequency range $\omega_\mathrm{q}/2\pi\simeq3$--$6~\mathrm{GHz}$. Time-domain measurements allow us to separate relaxation and dephasing contributions to the charge-qubit linewidth and reveal a crossover from dephasing-dominated decoherence at lower qubit frequencies to relaxation-dominated decoherence at higher frequencies. The relaxation rate follows a cubic dependence on qubit frequency, providing frequency-resolved evidence that charge relaxation is dominated by emission into the piezoelectric acoustic phonon bath in GaAs. These results demonstrate how tunable hybrid cQED devices can be used to identify microscopic decoherence channels in semiconductor quantum circuits.

\begin{figure*}[!t]
\includegraphics[width=\textwidth]{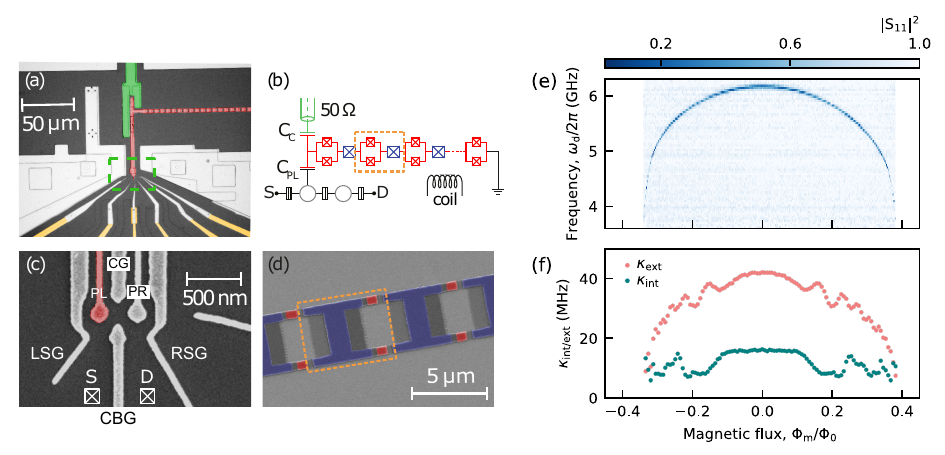}
\caption{
(a) False colored optical micrograph of the hybrid device. 
Al structures (light gray) are deposited on a GaAs/AlGaAs heterostructure (dark gray), to define the superconducting ground plane, and the fine gate structures utilized to form the double quantum dot (DQD). 
Au leads (yellow) connect the fine gates to the bonding pads (not shown in the figure).
The Al/AlOx/Al SQUID array resonator (red) is galvanically connected to one of the fine gates, which is driven with the $50~\Omega$ microwave feedline (green).
(b) Schematic circuit diagram displaying the DQD source ($S$), and drain ($D$) ohmic contacts, and coupling capacitance to the resonator ($C_{\rm{PL}}$). 
$C_\mathrm{c}$ denotes the capacitance between the resonator and the microwave feedline. 
Boxes with crosses indicate Josephson junctions (JJs) forming the SQUID array resonator. 
A superconducting coil is mounted above the sample to control the magnetic flux $\Phi_\mathrm{m}$ threading the SQUIDs.
The dashed orange box indicates a unit-cell of the SQUID array resonator. 
(c) False colored scanning electron micrograph of the gate structures defining the DQD in the green dashed box region in (a).
$S$ ($D$) denotes the source (drain) ohmic contact. 
The gate colored in red is galvanically connected to the SQUID array resonator. 
(d) False-colored scanning electron micrograph of the Al/AlOx/Al SQUID array resonator. Red structures denote the JJs forming the SQUIDs. 
Blue structures correspond to the larger-area series JJs (see main text).
As in (b), dashed orange box indicates a unit-cell of the SQUID array resonator. 
(e) Normalized resonator reflectance $|S_\mathrm{11}|^2$ measured as a function of the resonator drive frequency $\omega_\mathrm{d}$ and $\Phi_\mathrm{m}$.
(f) External (internal) photon loss rate of the resonator $\kappa_\mathrm{ext}$ ($\kappa_\mathrm{int}$) as a function of $\Phi_\mathrm{m}$ illustrated by red (green) dots.
$\kappa_\mathrm{ext}$ and $\kappa_\mathrm{int}$ are extracted from the circle-fit to the measured $|S_\mathrm{11}|^2$ in (e)}
\label{fig:SampleAndCircuit}
\end{figure*} 

\begin{figure*}[!t]
\includegraphics[width=\textwidth]
{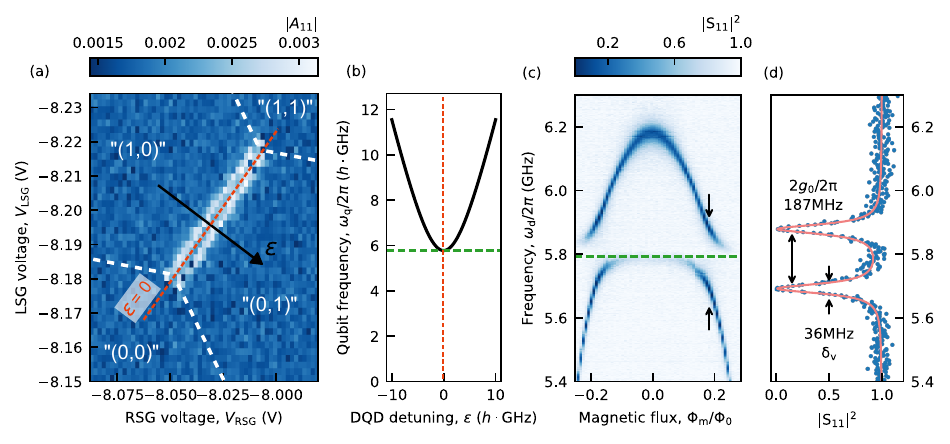}
\caption{
(a) Charge stability diagram measurement. The resonator reflection magnitude $|A_\mathrm{11}|$ is measured near an inter-dot transition, as a function of the voltages applied to the side barrier gates LSG ($V_\mathrm{LSG}$) and RSG ($V_\mathrm{RSG}$) (see Fig.~\ref{fig:SampleAndCircuit}(c)).
Black arrow represents the gate voltage direction in which the DQD detuning $\varepsilon$ is varied.
(b) DQD charge qubit frequency as a function of $\varepsilon$, calculated from $\omega_\mathrm{q}/2\pi = \sqrt{\varepsilon^2 + 4t^2}/h$ assuming the inter-dot tunnel coupling $t/h \sim 2.895~\mathrm{GHz}$. 
(c) $|S_\mathrm{11}|^2$ measured as a function of the resonator drive frequency $\omega_\mathrm{d}$ and the magnetic flux $\Phi_\mathrm{m}$, with $\omega_\mathrm{q}/2\pi$ tuned to $\sim 5.79$~GHz (green dashed line) by setting the gate voltages to maintain $\varepsilon = 0$ (yellow dashed line in (a) and (b)).
(d) Line-cut of the measured $|S_\mathrm{11}|^2$ in (c) along $\omega_\mathrm{d}$ at $\Phi_\mathrm{m}/\Phi_0 \sim 0.2$ (red arrows in (c)) revealing a vacuum-Rabi mode splitting.
The red solid line denotes the numerical fit to the master equation model, from which we extract the intrinsic charge-photon coupling strength $g_0/2\pi \sim 93.5$~MHz and the mode linewidth $\delta_\nu \sim 36$~MHz.
}
\label{fig:ResDQDChar}
\end{figure*}


To systematically study a charge qubit \cite{Petersson2010, hayashi2003, kim2015} strongly coupled to microwave photons, we employ a hybrid circuit quantum electrodynamics (circuit QED) architecture consisting of a gate-defined DQD capacitively coupled to a tunable, high-impedance superconducting resonator fabricated on a GaAs/AlGaAs heterostructure \cite{Stockklauser2017, Scarlino2019, scarlino2022}.
Fig.~\ref{fig:SampleAndCircuit}(a) shows a false-colored optical micrograph of a device identical to the one used in this work; the corresponding circuit schematic is shown in Fig.~\ref{fig:SampleAndCircuit}(b).
A scanning electron microscope (SEM) image of the gate-defined DQD, acquired from the region indicated by the green-dashed box in Fig.~\ref{fig:SampleAndCircuit}(a), is shown in Fig.~\ref{fig:SampleAndCircuit}(c).
The heterostructure hosts a high-mobility two-dimensional electron gas (2DEG) located $\approx 90$~nm below the surface, where a Si delta doping supplies carriers to the 2DEG.
By operating the device in depletion mode, we form a DQD under the plunger gates (see PL and PR Fig.~\ref{fig:SampleAndCircuit}(c)), which defines a dipolar charge qubit.
For the initial tuning procedure, the direct current (DC) through the source and drain ohmic contacts (S and D in Fig.~\ref{fig:SampleAndCircuit}(b, c)) is measured as a function of the gate voltages.

Importantly, to facilitate capacitive coupling between the DQD charge qubit and photons in the superconducting resonator, the gate shaded in red (Fig.~\ref{fig:SampleAndCircuit}(a, c)) is galvanically connected to a high-impedance superconducting resonator formed by a superconducting quantum interference device (SQUID) array \cite{Stockklauser2017}.
Fig.~\ref{fig:SampleAndCircuit}(d) shows the corresponding SEM image of a portion of the SQUID array, where Al/AlOx/Al Josephson junctions (JJs) (red shaded structures) are defined using the Dolan-bridge technique \cite{frunzio2005}.
We note that the blue shaded structures (Fig.~\ref{fig:SampleAndCircuit}(d)) correspond to spurious JJs also contributing to the total Josephson inductance of the SQUID array, as depicted in the circuit schematic (Fig.~\ref{fig:SampleAndCircuit}(b)).
While these spurious JJs are inevitably formed in Dolan-bridge-based SQUID arrays, the Josephson inductance is inversely proportional to the junction area of the corresponding JJ, implying that the inductances of the blue JJs are $\approx 10$ times smaller than those of the red JJs (Fig.~\ref{fig:SampleAndCircuit}(b, d), see Appendix~\ref{app:resonator_model}).
As shown in the schematic (Fig.~\ref{fig:SampleAndCircuit}(b)), the other end of the SQUID array is shunted to ground to form a distributed $\lambda$/4 resonator \cite{Blais2021}.
The resonator exhibits a voltage antinode near the DQD gate, which maximizes the charge-photon coupling strength.
A 50~$\Omega$ microwave feedline (green structure in Fig.~\ref{fig:SampleAndCircuit}(a, b)) is capacitively coupled to the resonator near the voltage antinode to operate in reflection mode.

To enable broad tuning of the resonance frequency of the SQUID array resonator $\omega_\mathrm{r}$, we mount a superconducting coil above the device in the sample package (not shown in the figure) to control the magnetic flux $\Phi_\mathrm{m}$ that threads each of the SQUID loops.
$\Phi_\mathrm{m}$ tunes the Josephson inductance of each SQUID, and thereby the total effective inductance of the resonator $L_\mathrm{tot}$.
Because $\omega_\mathrm{r}/2\pi = 1/\sqrt{L_\mathrm{tot}(\Phi_\mathrm{m}) C_\mathrm{tot}}$, with $C_\mathrm{tot}$ denoting the total effective capacitance of the resonator (see Appendix~\ref{app:resonator_model} for details), $\Phi_\mathrm{m}$ can be used to adjust $\omega_\mathrm{r}$.
Fig.~\ref{fig:SampleAndCircuit}(e) presents the normalized resonator reflectance $|\mathrm{S}_{11}|^2 = |A_{11}/A_{0}|^2$ as a function of the applied normalized flux $\Phi_\mathrm{m}/\Phi_0$ and feedline drive frequency $\omega_\mathrm{d}$, with the charge qubit frequency far detuned from the resonance frequency of the resonator.
\add{Here, $A_{11}$ is the resonator reflectance, $A_0$ is the background signal of the measurement chain, $\Phi_0=h/2e$ is the magnetic flux quantum\delete{, with Planck constant $h$ and electron charge $e$}.}
At $\Phi_\mathrm{m}/\Phi_0 = 0$, the total inductance and capacitance are estimated to be $L_\mathrm{tot} \approx 30.6$~nH, including the inductance of spurious JJs, and $C_\mathrm{tot} \approx 20$~fF, respectively. These values yield a resonance frequency of $\omega_\mathrm{r}/2\pi \approx 6.2$~GHz, which can be tuned to below 4~GHz at finite $\Phi_\mathrm{m}$ (see Appendix~\ref{app:flux_beta_kext}).
By fitting each line cut of Fig.~\ref{fig:SampleAndCircuit}(e) along $\omega_\mathrm{d}$ using a circle-fit procedure, we obtain the feedline-resonator coupling strength $\kappa_\mathrm{ext}$ and the internal resonator loss rate $\kappa_\mathrm{int}$ as reported in Fig.~\ref{fig:SampleAndCircuit}(f). 
This indicates that the device is operated in the overcoupled regime with $\kappa_\mathrm{ext}/\kappa_\mathrm{int} > 1$ \cite{Blais2021} for most of the measured frequency range, allowing to achieve a high signal-to-noise ratio (SNR) in the reflection signal.

By measuring $|\mathrm{S}_{11}|^2$ as a function of the gate voltages $V_{\rm LSG}$ and $V_{\rm RSG}$, we obtain a charge stability diagram of a representative DQD configuration near an inter-dot charge transition as shown in Fig.~\ref{fig:ResDQDChar}(a) (see Appendix~\ref{app:large_csd} for a stability diagram measured in a larger gate voltage range).
``($n_\mathrm{L}$, $n_\mathrm{R}$)" denote the effective charge occupancy of the DQD only taking into account the number of excess charges in the DQD, with $n_\mathrm{L}$ ($n_\mathrm{R}$) indicating the number of electrons in the left (right) QD. 
A dipolar charge qubit can be naturally defined using the DQD, whose dynamics is described by the Hamiltonian $H_\mathrm{cq} = (\varepsilon/2) \,\tau_z + t \, \tau_x$.
Here, $\varepsilon = \mu_\mathrm{L} - \mu_\mathrm{R}$ is the DQD detuning, with $\mu_\mathrm{L}$ ($\mu_\mathrm{R}$) denoting the chemical potential of the left (right) QD, $t$ is the inter-QD tunnel coupling strength, and $\tau_i$  is the Pauli-$i$ operator in the charge basis [``(1,0)", ``(0,1)"] \cite{vanderwiel2002}.
The Hamiltonian results in the charge qubit frequency $\omega_{\rm q}/2\pi = \sqrt{4t^2 + \varepsilon^2}/h$, which is illustrated in Fig.~\ref{fig:ResDQDChar}(b) as a function of $\varepsilon$.
Notably, the device geometry supports a tunability over a wide range of $t$ which is central for studying the charge coherence as we discuss below.

As can be inferred from the schematic spectrum of $\omega_{\rm q}$ shown in Fig.~\ref{fig:ResDQDChar}(b), $\varepsilon = 0$ corresponds to a charge coherence sweet spot with $\partial \omega_\mathrm{q} / \partial \varepsilon = 0$, which implies first-order protection against detuning noise for the charge qubit \cite{Paladino2014, Petersson2010, hayashi2003}.
To estimate the value of $2t$, we first fix the gate voltages close to the DQD detuning sweet spot $\varepsilon = 0$, such that the corresponding charge qubit frequency satisfies $\omega_\mathrm{q}/2\pi \sim 2t/h$. 
Then, the magnetic flux $\Phi_\mathrm{m}$ is varied to tune the fundamental mode of the resonator across the charge qubit transition, and $|S_{11}|^2$ is recorded, as shown in Fig.~\ref{fig:ResDQDChar}(c).
In contrast to the similar spectroscopy data presented in Fig.~\ref{fig:SampleAndCircuit}(e), where the qubit frequency $\omega_\mathrm{q}$ was far detuned from $\omega_\mathrm{r}$, the measured spectrum in Fig.~\ref{fig:ResDQDChar}(c) exhibits clear avoided crossings, when $\omega_\mathrm{r}/2\pi = \omega_\mathrm{q}/2\pi \sim 2t/h \sim 5.8$~GHz.

The linecut along $\omega_\mathrm{d}$ at $\Phi_\mathrm{m}/\Phi_0 \sim 0.2$ (black arrows in Fig.~\ref{fig:ResDQDChar}(c)), where the resonator mode is resonant with the qubit transition ($\omega_\mathrm{r} = \omega_\mathrm{q}$), is shown in Fig.~\ref{fig:ResDQDChar}(d). 
The solid red line is a fit of the measured spectrum to a master-equation model (see Appendix~\ref{app:master_equation}), from which we extract the qubit decoherence rate $\Gamma_{2}/2\pi \sim 11.81~\mathrm{MHz}$ and a coupling strength $2g_0/2\pi \sim 187 \pm 5~\mathrm{MHz}$, with the total linewidth of the bare resonator $\kappa = \kappa_\mathrm{ext} + \kappa_\mathrm{int} \sim 49.37$~MHz at $\omega_\mathrm{r}/2\pi \sim 5.8$~GHz (see Fig.~\ref{fig:SampleAndCircuit}(f)).
The linewidth of the vacuum-Rabi doublet at resonance is given by $\delta_\nu = \Gamma_2 + \kappa/2 \sim 36$~MHz.
This places the device in the strong-coupling regime, with $g_0/2\pi\simeq93.5~\mathrm{MHz}$ exceeding both $\kappa/2\pi\simeq49.4~\mathrm{MHz}$ and $\Gamma_2/2\pi\simeq11.8~\mathrm{MHz}$.

\add{
Notably, the near-unity visibility of the vacuum-Rabi doublet arises from an effective impedance matching between the external coupling and the dressed decay channels of the hybridized resonator--qubit system. 
At resonance, the reflection at each vacuum-Rabi mode frequency $\omega_\pm$ is governed by the balance between external coupling and the effective internal loss of the dressed modes, $|S_{11}(\omega_\pm)| \propto (\kappa_{\mathrm{ext}} - \kappa_{\pm,\mathrm{int}})/(\kappa_{\mathrm{ext}} + \kappa_{\pm,\mathrm{int}})$, where $\kappa_{\pm,\mathrm{int}} = \kappa_{\mathrm{int}} + 2\,\mathrm{Im}[g_{\mathrm{eff,res}}\chi(\omega_\pm)]$, and $\chi(\omega) = g_{\mathrm{eff,res}}/(\omega_q - \omega - i\Gamma_2)$ denotes the qubit susceptibility (see Appendix~\ref{app:master_equation}). 
Here, $g_\mathrm{eff,res}$ denotes the effective charge-photon coupling strength which also takes into account the reduced charge dipole moment away from the sweet spot (see Appendix~\ref{app:g0_scaling}).
In our device, the condition $\kappa_{\mathrm{ext}} \approx \kappa_{\mathrm{int}} + 2\,\mathrm{Im}[g_{\mathrm{eff,res}}\chi(\omega_\pm)]$ places the system close to critical coupling for each vacuum-Rabi mode, thereby maximizing the visibility of the vacuum-Rabi splitting.
}

\begin{figure}[h]
\includegraphics[width=\columnwidth]{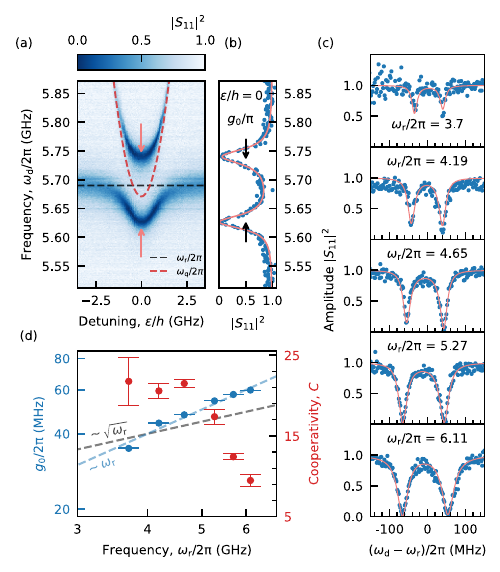}
\caption{
(a) $|S_\mathrm{11}|^2$ measured as a function of $\omega_\mathrm{d}$ and $\varepsilon$ with $\omega_\mathrm{r}/2\pi \sim 2t/h \sim  5.69$~GHz.
Dashed black line corresponds to the calculated $\omega_\mathrm{q}/2\pi$, and the dashed red line denotes $\omega_\mathrm{r}/2\pi$, using the parameters extracted from the 2D fit of the spectrum to a master equation model (see Appendix~\ref{app:master_equation}).
(b) Linecut of the measured $|S_\mathrm{11}|^2$ in (a) at $\varepsilon = 0$ (red arrows) revealing the vacuum-Rabi mode splitting. 
The red solid line is the numerical fit to a master equation model (see text)
(c) $|S_\mathrm{11}|^2$ measured as a function of $\omega_\mathrm{d}$ at $\varepsilon = 0$ with different values of $t/h$, achieved by gate voltage tuning while maintaining the constant number of charges. 
$\omega_\mathrm{r}$ is accordingly tuned with $\Phi_\mathrm{m}$ (see Fig.~\ref{fig:SampleAndCircuit}(e)), to be resonant with the qubit at $\varepsilon = 0$ i.e. $\omega_\mathrm{r}/2\pi \sim 2t/h$. 
The red solid line in each panel is the numerical fit to the master equation model (see Table~\ref{tab1:para_fig3} for the parameters extracted from the fit in the Appendix~\ref{app:hybrid_spectrum_different_2t}).
See Appendix Fig.~\ref{fig:supplementary_gvsfreq} for the full measurement of the resonator spectrum as a function of $\varepsilon$.
(d) A log-log plot of $g_0$ values extracted from the fits of the measured $|S_\mathrm{11}|^2$ to the master equation model in (c), as a function of $\omega_\mathrm{r}$ (blue dots). 
Corresponding cooperativity values $C = g_0^2/(\kappa\Gamma_2)$ (red dots) are calculated also using the parameters extracted from the fits.
While $g_0$ is expected to be proportional to $\sqrt{\omega_\mathrm{r}}$ (black dashed line), a proportionality to $\omega_\mathrm{r}$ (blue dashed line) is observed (see text).
}
\label{fig:Resonator_Response}
\end{figure}

Having established strong charge--photon coupling using resonator-frequency tuning at fixed \replace{DQD electrostatics}{DQD detuning $\varepsilon$ and tunnel coupling $t$}, we next investigate the coherence properties of the hybrid system by studying the DQD--resonator hybridization as a function of DQD detuning for several values of the inter-dot tunnel coupling $t$. For each tunnel-coupling configuration, the resonator frequency is tuned such that the resonant condition $\omega_{\rm r}/2\pi=\omega_{\rm q}(\varepsilon=0)/2\pi = 2t/h$ is satisfied at the sweet spot ($\varepsilon=0$).
Importantly, measurements of the vacuum-Rabi splitting at $\varepsilon = 0$ provide direct access to the intrinsic charge--photon coupling strength $g_0$, independent of detuning-induced mixing of charge states (see Appendix~\ref{app:g0_scaling}). 
The measurements presented here are performed in a DQD configuration with a different charge occupation in each QD, and with a different tunnel coupling $t$ compared to Fig.~\ref{fig:ResDQDChar}, due to an unexpected shift in the DQD gate voltages.
To accurately reach the resonant condition at $\varepsilon = 0$ ($\omega_{\rm r}/2\pi \sim 2t/h \sim 5.69$~GHz) as the one reported in Fig.~\ref{fig:Resonator_Response}(a) for an initially unknown $2t$, we measure the resonator response $|S_{11}|^2$ as a function of both $\varepsilon$ and the probe frequency $\omega_\mathrm{d}$ for several values of $\omega_{\rm r}$, as shown in Fig.~\ref{fig:supplementary_gvsfreq} in Appendix~\ref{app:hybrid_spectrum_different_2t}. 
The measured two-dimensional spectrum corresponding to the resonant configuration, shown in Fig.~\ref{fig:Resonator_Response}(a), is fitted to a master-equation model (Appendix~\ref{app:master_equation}). This procedure yields an intrinsic coupling strength of $g_0/2\pi \sim 57.43 \pm 0.02$~MHz, a resonator frequency $\omega_\mathrm{r}/2\pi \sim 5.69$~GHz (black dashed line in Fig.~\ref{fig:Resonator_Response}(a)), a tunnel coupling $2t/h \sim 5.671$~GHz, and a total decoherence rate $\Gamma_2=\Gamma_\phi+\Gamma_1/2$ of the charge qubit (see Table~\ref{tab1:para_fig3} for additional extracted parameters and Appendix~\ref{app:hybrid_spectrum_different_2t} Fig.~\ref{fig:supplementary_gvsfreq} for 2D fits of all measured configurations). 
Here, $\Gamma_\phi$ and $\Gamma_1$ denote the pure dephasing and relaxation rates, respectively, whose frequency dependence will be analyzed in detail in the following sections.
The fitting procedure also enables precise identification of the DQD detuning sweet spot ($\varepsilon = 0$), indicated by the red arrows in Fig.~\ref{fig:Resonator_Response}(a). A line cut of the $S_{11}$ at $\varepsilon = 0$ is shown in Fig.~\ref{fig:Resonator_Response}(b), clearly resolving the vacuum-Rabi doublet.
The solid red line superposed on the measured vacuum-Rabi doublet is a fit to the master equation model, obtained during the two-dimensional fitting procedure.
Using the extracted tunnel coupling $2t/h$, we calculate the charge-qubit transition frequency $\omega_\mathrm{q}/2\pi=\sqrt{(2t/h)^2+(\varepsilon/h)^2}$, represented by the red dashed curve in Fig.~\ref{fig:Resonator_Response}(a). 

Fully exploiting the tunability of both the SQUID-array resonator and the DQD electrostatic potential, we investigate the intrinsic charge--photon coupling strength $g_0$ for several values of the inter-dot tunnel coupling $t$, while keeping the number of charges in the DQD fixed. 
For each value of $t$, the resonator flux $\Phi_{\rm m}$ and the DQD gate voltages are iteratively adjusted such that the resonant condition $\omega_{\rm r}/2\pi \simeq \omega_{\rm q}(\varepsilon=0)/2\pi = 2t/h$
is satisfied at the DQD detuning sweet spot. 
The linecuts at $\varepsilon = 0$ for the other values of the studied $2t$ are reported in Fig.~\ref{fig:Resonator_Response}(c), with the corresponding fits to the master-equation model described in Appendix~\ref{app:master_equation}; see solid red lines in Fig.~\ref{fig:Resonator_Response}(c).

\replace{The intrinsic coupling strengths $g_0$ extracted from the fits for six different values of $2t$ are reported in Fig.~\ref{fig:Resonator_Response}(d) (blue circles).}{The intrinsic coupling strengths $g_0$ extracted from the fits for six different values of $2t$ are reported as blue circles in the log-log plot shown in Fig.~\ref{fig:Resonator_Response}(d).} 
For an ideal inductively tuned resonator with constant effective capacitance $C_{\rm eff}$ and fixed DQD detuning lever arm, one expects a scaling $g_0\propto\sqrt{\omega_{\rm r}}$ (dashed black line in Fig.~\ref{fig:Resonator_Response}(d)) (see Appendix~\ref{app:g0_scaling}). Instead, the measured values clearly deviate from this dependence and are more consistent with an approximately linear scaling $g_0\propto\omega_{\rm r}$ (dashed blue line).

In general, the coupling strength is determined by the voltage fluctuations at the coupling node, $g_0\propto\alpha V_{\rm zpf}$, where $V_{\rm zpf}=\sqrt{\hbar\omega_{\rm r}/(2C_{\rm eff})}$ is the resonator zero-point voltage, or equivalently $g_0\propto\alpha\,\omega_{\rm r}\sqrt{Z_{\rm r}}$ in terms of the mode impedance $Z_{\rm r}$. Here $\alpha$ denotes the DQD detuning lever arm of the gate connected to the resonator, which depends on the effective electric dipole moment of the charge qubit \cite{scarlino2022}. 
The observed linear scaling therefore indicates that either the effective impedance $Z_{\rm r}$ does not follow the expected dependence on the resonator frequency, $Z_{\rm r} \propto 1/\omega_\mathrm{r}$ (see Appendix~\ref{app:g0_scaling}), or that the effective dipole moment of the charge qubit changes as the tunnel coupling $t$ is varied.

To distinguish between these two possibilities, we perform a complementary set of measurements in which the tunnel coupling $t$ and the entire DQD electrostatic configuration are kept fixed, while only the resonator frequency $\omega_{\rm r}$ is tuned by varying the applied magnetic flux $\Phi_{\rm m}$ (see Appendix~\ref{app:hybrid_spectrum_different_w_r}). In this configuration, the resonance condition $\omega_{\rm r}=\omega_{\rm q}$ is reached at finite DQD detuning $\varepsilon\neq0$, and the observed avoided crossing reflects a reduced effective coupling 
$g_{\rm eff}=g_0\left(\frac{2t}{\hbar\omega_{\rm r}}\right)$
arising from the reduction of the charge-qubit dipole moment at finite $\varepsilon$ (see Appendix~\ref{app:g0_scaling}). After correcting for this mixing angle, the intrinsic coupling strength $g_0$ is extracted for each resonator frequency (Appendix Fig.~\ref{fig:supplementary_scaling}(c)). 

Remarkably, the extracted values of $g_0$ again exhibit the same approximately linear dependence on $\omega_{\rm r}$ as observed in the primary measurement  (see Fig.~\ref{fig:supplementary_scaling} in Appendix~\ref{app:hybrid_spectrum_different_w_r}). 
This supports the interpretation that the observed scaling is mainly associated with the electromagnetic properties of the distributed high-impedance resonator, including flux-dependent impedance, capacitive loading, and mode-shape, rather than being solely due to changes in the DQD dipole moment.

Furthermore, using the parameters extracted from the master-equation fit in Fig.~\ref{fig:Resonator_Response}(c), we calculate the corresponding cooperativity $C = g_0^2/(\kappa \Gamma_2)$, a figure of merit that quantifies the coherence of a light--matter hybrid interface (red circles in Fig.~\ref{fig:Resonator_Response}(d)) \cite{cottet2017} .
Despite the approximately linear increase of $g_0$ with $\omega_\mathrm{r}$, the cooperativity $C$ is maximized at lower resonator frequencies, when $\omega_\mathrm{r} = \omega_\mathrm{q}$, reaching $C \approx 20$ at $\omega_\mathrm{q}/2\pi = 3.71$~GHz.
This value ranks among the highest cooperativities reported for resonator--DQD charge hybrid systems \cite{cottet2017, janik2024}.
Such a high cooperativity $C$ primarily originates from a reduced $\Gamma_2$ at small tunnel coupling $t$ of the DQD, which we investigate in detail in the following.

While numerous experiments have established low-frequency charge noise as the primary limitation to dephasing of semiconductor charge qubits~\cite{Thorgrimsson2017,Petersson2010, kim2015}, the microscopic origin of relaxation in these qubits remains less directly understood.
Electron--phonon coupling has long been identified as a dominant relaxation channel in DQDs~\cite{Brandes2005, Hofmann2020, golovach2004}.
In particular, in GaAs, which lacks inversion symmetry, electrons couple to acoustic phonons via both deformation-potential and piezoelectric mechanisms.
The piezoelectric interaction arises from strain-induced electric fields generated by lattice vibrations, whereas DP coupling originates from band-structure modulation under lattice compression~\cite{Brandes2005}.
In the GHz regime relevant for DQD charge qubits in cQED architectures ($\omega_\mathrm{q}/2\pi \sim 0.1$--$10$~GHz), piezoelectric coupling is expected to dominate due to its larger low-frequency spectral weight~\cite{Brandes2005}.
Microscopically, the relaxation rate $\Gamma_1$ is governed by the environmental spectral density evaluated at the qubit transition frequency, $J(\omega_\mathrm{q})$, i.e., $\Gamma_1(\omega_\mathrm{q}) \propto J(\omega_\mathrm{q})$.
For piezoelectric coupling in a three-dimensional crystal, the combined effects of the phonon density of states and the charge-qubit matrix element lead to the characteristic cubic scaling $\Gamma_1 \propto \omega_\mathrm{q}^3$ at $\varepsilon = 0$ (see Appendix~\ref{app:charge_qubit_relaxation_theory} and Ref.~\cite{Hofmann2020}).
Measuring $\Gamma_1$ as a function of qubit frequency therefore provides a direct probe of $J(\omega_\mathrm{q})$.

To independently extract the relaxation ($\Gamma_1$) and pure-dephasing ($\Gamma_\phi$) rates of the charge qubit, we perform time-resolved measurements~\cite{Scarlino2019}.
Throughout the manuscript, decay rates plotted or tabulated in MHz are reported as angular rates divided by $2\pi$, i.e. $\Gamma_i/2\pi=1/(2\pi T_i)$.
Exploiting the frequency tunability of the SQUID-array resonator, we maintain the system deeply in the dispersive regime ($\Delta_{\rm r,q} = \omega_{\rm r} - \omega_{\rm q} \geq 15 g_0$) for all the studied configurations with different values of $2t$, (Appendix Table~\ref{tab1:para_fig4}), thereby ensuring a high signal-to-noise ratio across the full tuning range.
In contrast, for a fixed-frequency resonator, the signal would rapidly diminish at large frequency detuning due to the reduced dispersive shift $\chi = g_0^2/\Delta_{\rm r,q}$.
The tunability of our SQUID array resonator is therefore essential for extracting the decoherence rates at different charge qubit energies.

\begin{figure}[t]
\includegraphics[width=\columnwidth]{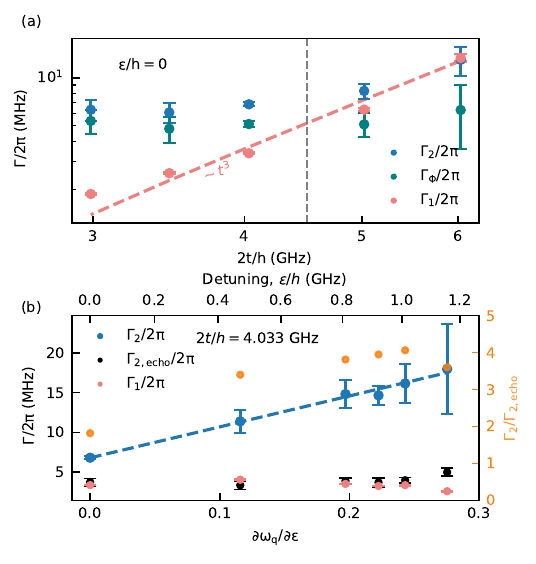}
\caption{
(a) Relaxation ($\Gamma_1$) and Ramsey decay ($\Gamma_2$) rates of the DQD charge qubit extracted from time-resolved measurements (see Appendix~\ref{app:time_resolved_measurements}) vs. tunnel coupling $t$, while maintaining the DQD detuning at $\varepsilon = 0$.
Number of charges in each QD are kept constant throughout the measurements.
Pure dephasing rates ($\Gamma_\phi$) are calculated from the relation $\Gamma_\phi = \Gamma_2 - \Gamma_1/2$.
The red dashed line is the fit of the extracted $\Gamma_1$ to Eq.~(\ref{eq:fitT1}).
(b) Ramsey ($\Gamma_2$), Relaxation ($\Gamma_1$) and Hahn-echo ($\Gamma_\mathrm{2,echo}$) decay rates as a function of $\partial \omega_\mathrm{q}/\partial \varepsilon$ extracted at finite $\varepsilon$ (top axis) with a fixed $2t/h = 4.033$~GHz.
The blue dashed line is a fit of $\Gamma_2$ to Eq.~(\ref{eq:sigma_noise}).
The orange points illustrate $\Gamma_2/\Gamma_\mathrm{2,echo}$ ratio as a function of $\partial \omega_\mathrm{q}/\partial \varepsilon$.
}
\label{fig:Parameters}
\end{figure}

Representative free-induction decay (FID, i.e., Ramsey), Hahn-echo, and charge qubit relaxation measurements are presented in Appendix Fig.~\ref{fig:supplementary_timedomain}, together with the corresponding time traces of the dispersive readout signal~\cite{Scarlino2019}.
We repeat these measurements for different values of $t$ and $\varepsilon$, and extract the corresponding coherence times $T_\mathrm{2,Ramsey}$, $T_\mathrm{2,echo}$, and $T_1$ by fitting to exponential decay models \cite{Scarlino2019}.
The extracted relaxation ($\Gamma_1 = 1/T_1$) and Ramsey decay ($\Gamma_2 = 1/T_\mathrm{2,Ramsey}$) rates at $\varepsilon = 0$ are shown in Fig.~\ref{fig:Parameters}(a), together with the inferred pure dephasing rate $\Gamma_\phi = \Gamma_2 - \Gamma_1/2$, which accounts for inhomogeneous dephasing processes~\cite{Schuster2005}.
For the lowest qubit frequencies ($\omega_\mathrm{q}/2\pi < 4.2$~GHz), we obtain $g_0/\Gamma_2 \sim 10$ and a cooperativity of $C \sim 20$ (Appendix Table~\ref{tab1:para_fig3}).
Accessing even lower qubit frequencies would further enhance $C$, but lies outside the 4--8~GHz bandwidth of the readout-chain of the measurement setup.

Notably, the extracted $\Gamma_1$ at $\varepsilon = 0$ for different values of $t$ (red points in Fig.~\ref{fig:Parameters}(a)) follows the expected cubic dependence $\Gamma_1 \propto \omega_\mathrm{q}^3$  (see Appendix~\ref{app:charge_qubit_relaxation_theory}), and is well described by
\begin{equation}
\Gamma_1^{\rm phn}\big[\omega_{\rm q}/2\pi(\varepsilon = 0)\big] \sim a(\omega_{\rm q}/2\pi)^3 + b,
\label{eq:fitT1}
\end{equation}
where $a$ quantifies the strength of the piezoelectric (PE) coupling, and $b$ captures additional frequency-independent relaxation channels.
The observed cubic scaling thus provides direct evidence that relaxation is dominated by the emission of piezoelectric acoustic phonons in the explored frequency range.

The time-domain measurements further reveal a crossover in the dominant decoherence mechanism at the detuning noise sweet spot.
For $2t/h\lesssim4.5~\mathrm{GHz}$ (vertical dashed gray line in Fig.~\ref{fig:Parameters}(a)), the qubit decoherence rate $\Gamma_2$ is dominated by pure dephasing $\Gamma_\phi$. 
At higher qubit frequencies, the relaxation contribution increases rapidly and becomes comparable to, and eventually larger than, the pure-dephasing contribution, indicating a crossover toward relaxation-dominated decoherence.
In contrast to $\Gamma_1$, the extracted $\Gamma_\phi$ exhibits only a weak dependence on $t$.
\add{
The reduction of $\Gamma_2$ due to the suppressed $\Gamma_1$ at lower $\omega_\mathrm{q}$ naturally enhances the cooperativity $C = g_0^2/(\kappa \Gamma_2)$, explaining the increase of $C$ at lower $\omega_\mathrm{q}$ observed in Fig.~\ref{fig:Resonator_Response}(d) despite $g_0 \propto \omega_\mathrm{r}$.
These results indicate that reducing $t$ may be advantageous for charge-coherence optimization when relaxation constitutes the dominant decoherence channel. 
}

We further investigate the qubit coherence away from the detuning noise sweet spot in the regime where $\Gamma_2$ dominates over $\Gamma_1$, while fixing $2t/h \sim 4.033$~GHz.
$\Gamma_2$ increases proportionally with $\partial \omega_\mathrm{q}/\partial \varepsilon$, see Fig.~\ref{fig:Parameters}(b)~\cite{Thorgrimsson2017,Ithier2005}.
For quasistatic Gaussian detuning fluctuations with standard deviation $\sigma_{\varepsilon}$, the Ramsey decay rate follows
\begin{equation}
\Gamma_2 = \frac{\left|\partial \omega_{\rm q}/\partial \varepsilon\right| \, \sigma_{\varepsilon}}{\sqrt{2}\hbar}.
\label{eq:sigma_noise}
\end{equation}
\clearpage

Fitting the extracted $\Gamma_2$ to Eq.~\ref{eq:sigma_noise} yields $\sigma_{\varepsilon} \sim 0.23 \pm 0.03~\mu\text{eV}$, consistent with previous observations in similar architectures~\cite{Scarlino2019}.
The increase of $\Gamma_2$ at finite $\varepsilon$ is therefore primarily attributed to increased dephasing, as supported by the extracted $\Gamma_1$, which remains nearly constant (or slightly decreases) as a function of $\varepsilon$ (Fig.~\ref{fig:Parameters}(b)).
Similar trends have been reported in SiGe DQDs~\cite{Wang2013c,Kim2015a} and in Cooper-pair-box qubits~\cite{Astafiev2004}.
\add{
Moreover, as evident from Eq.~\eqref{eq:sigma_noise}, charge-detuning noise is suppressed to first order at the sweet spot ($\varepsilon = 0$). 
As a consequence, second-order dephasing processes may become dominant, potentially leading to a polynomial decay of the FID profile, which exhibits a smoother decay than an exponential profile \cite{ramon2022qubit, nichol2017high}.
}

Furthermore, $\Gamma_\mathrm{2,echo} = 1/T_\mathrm{2,echo}$ as a function of $\varepsilon$ is shown in Fig.~\ref{fig:Parameters}(b) (black points).
\add{
$\Gamma_\mathrm{2,echo}$ does not show an explicit dependence on $\partial \omega_\mathrm{q} / \partial \varepsilon$ (or $\varepsilon$), and remains constant ($\Gamma_\mathrm{2,echo}/2\pi \sim 4$~MHz) over the measured range, which clearly demonstrates that the echo effectively cancels out the low-frequency DQD detuning noise in the system.
}
As a result, the ratio $\Gamma_2 / \Gamma_\mathrm{2,echo}$ (orange points in Fig.~\ref{fig:Parameters}(b)) increases from $\sim 1.9$ to $\sim 4$ with increasing $\varepsilon$, indicating that the echo sequence suppresses a larger fraction of low-frequency charge noise at larger DQD detuning.
This observation further supports that $\Gamma_2$ at finite $\varepsilon$ is dominated by dephasing rather than relaxation. 
While relaxation is governed by noise at the qubit frequency and dephasing is sensitive to low-frequency noise~\cite{Ithier2005}, the echo sequence is only highly effective at suppressing the linear low-frequency fluctuations associated with dephasing.

In summary, we explored a hybrid circuit QED platform consisting of a frequency-tunable high-impedance SQUID-array resonator coupled to a GaAs double quantum dot charge qubit. The wide tunability enables time-domain measurements over a broad range of qubit frequencies while maintaining high signal-to-noise ratio dispersive readout.
By tuning the qubit frequency, we distinguish relaxation and dephasing contributions to decoherence and observe a crossover from dephasing- to relaxation-dominated decoherence near the charge sweet spot. The relaxation rate follows a cubic dependence on qubit frequency, consistent with piezoelectric acoustic phonon emission in GaAs.
More generally, these results demonstrate that tunable high-impedance resonators can serve as quantitative probes of microscopic dissipation mechanisms in semiconductor quantum devices.
This approach can be extended to other materials and hybrid architectures to identify and ultimately mitigate the dominant noise sources limiting coherence.

\textbf{Acknowledgments}
P.S., A.W., K.E., T.I., and S.B. acknowledge support from the NCCR SPIN, a National Centre of Competence in Research, funded by the Swiss National Science Foundation (SNSF) under grant number 225153.
P.S. acknowledges support from the Swiss State Secretariat for Education, Research and Innovation (SERI) under contract number MB22.00081. 
A.W. acknowledges support from NCCR Quantum Science and Technology, the project Elements for Quantum Information Processing with Semiconductor/Superconductor Hybrids (EQUIPS), and ETH Zurich.
V.F.M thanks NanoLund for financial support.
\appendix

\makeatletter
\twocolumngrid
\makeatother

\setcounter{secnumdepth}{2}

\renewcommand{\thesection}{A\arabic{section}}
\renewcommand{\thesubsection}{\thesection.\arabic{subsection}}
\renewcommand{\theequation}{\thesection.\arabic{equation}}

\makeatletter
\@addtoreset{equation}{section}
\makeatother

\setcounter{section}{0}


\section{Charge stability diagram measured in a broad gate voltage range}
\label{app:large_csd}

\begin{figure}[b]
\includegraphics[width=\columnwidth]{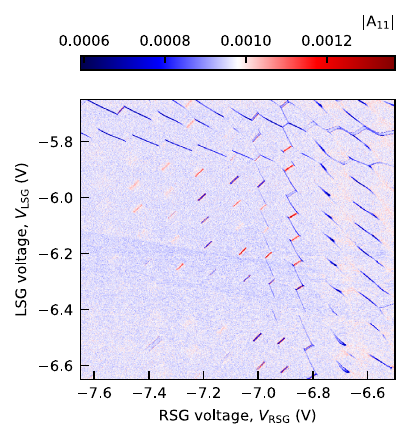}

\caption{
Resonator reflection $|A_{11}|$ measured as a function of the $V_\mathrm{LSG}$ and $V_\mathrm{RSG}$ (see Fig.~\ref{fig:SampleAndCircuit}(c)). The resonator frequency was tuned to $\omega_\mathrm{r}/2\pi \sim 5.9$~GHz by applying a finite magnetic flux $\Phi_\mathrm{m}$. 
}

\label{fig:supp_largeCSD}
\end{figure}

In this section, we present a charge stability diagram measured over a broader gate-voltage range than the one shown in Fig.~\ref{fig:ResDQDChar}(a), in order to further confirm the formation of the DQD. 
Fig.~\ref{fig:supp_largeCSD} shows the resonator reflection magnitude $|A_{11}|$ measured as a function of the gate voltages $V_\mathrm{LSG}$ and $V_\mathrm{RSG}$ (see Fig.~\ref{fig:SampleAndCircuit}(c)). 
For this measurement, the resonator frequency was tuned to $\omega_\mathrm{r}/2\pi \sim 5.9$~GHz by applying a finite magnetic flux $\Phi_\mathrm{m}$ (see Fig.~\ref{fig:SampleAndCircuit}(e)). 
The resonator probes interdot transitions both resonantly and dispersively when the corresponding transition frequencies are sufficiently close to $\omega_\mathrm{r}$.

\section{SQUID-array resonator: circuit model and parameters}
\label{app:resonator_model}

The Josephson junctions in the SQUID-array resonator were fabricated using the Dolan-bridge technique~\cite{Dolan1977}. 
The parameters of the array were chosen such that the fundamental mode at zero applied flux lies near $6~\mathrm{GHz}$, within the measurement bandwidth of the detection chain, while remaining well separated from the second resonant mode of the SQUID array, expected around $18$--$19~\mathrm{GHz}$~\cite{Masluk2012}. Figure~\ref{fig:SampleAndCircuit}(b,d) shows the circuit model of the SQUID-array resonator and a scanning electron micrograph of the array.

Each unit cell of the array consists of a SQUID formed by two nominally identical small Josephson junctions connected in parallel, in series with a larger-area Josephson junction. The SQUID junctions, shown in red in Fig.~\ref{fig:SampleAndCircuit}(d), have an inductance $L_{\mathrm{Sq}}^0=L_{\mathrm{Sq}}(\Phi_\mathrm{m}=0)\simeq1.25~\mathrm{nH}$ at zero flux and a capacitance $C_{\mathrm{Sq}}\simeq80~\mathrm{fF}$. 
The larger junctions, shown in blue in Fig.~\ref{fig:SampleAndCircuit}(d), are approximately ten times larger in area and have inductance $L_J^\star\simeq0.11~\mathrm{nH}$ and capacitance $C_J^\star\simeq800~\mathrm{fF}$\footnote{$C_{\mathrm{Sq}}$ and $C_J^\star$ were estimated from the junction areas using a capacitance density of $40~\mathrm{fF}/\mu\mathrm{m}^2$.}. 
Owing to their larger critical current, the large junctions contribute a smaller but non-negligible, approximately flux-independent inductance to the array.

From room-temperature resistance measurements, the total Josephson inductance of the SQUID junctions at zero flux is estimated to be
\begin{equation}
L_{\mathrm{r,Sq}}^0 = N L_{\mathrm{Sq}}^0/2 \simeq 20~\mathrm{nH},
\end{equation}
where $N=34$ is the number of unit cells. The total inductance associated with the larger in-series junctions is
\begin{equation}
L_{\mathrm{r,J}} = N L_J^\star \simeq 2~\mathrm{nH}.
\end{equation}
Additional flux-independent contributions arise from the geometric inductance, $L_g\simeq0.3~\mathrm{nH}$, estimated from electromagnetic simulations, and from the kinetic inductance of the aluminum wiring, $L_k\simeq1.5~\mathrm{nH}$~\cite{Annunziata2010}.

Each unit cell has an average stray capacitance to ground $C_0\simeq C_{\mathrm{gnd}}/N\simeq0.55~\mathrm{fF}$. The capacitive couplings of the resonator to the double quantum dot, to the surrounding metallic gates, and to the microwave feedline are denoted by $C_{\mathrm{PL}}$, $C_{\mathrm{gates}}$, and $C_c$, respectively.

Within an effective lumped-element approximation for the fundamental mode of the $\lambda/4$ resonator~\cite{Goeppl2009}, the flux-dependent resonance frequency and characteristic impedance are written as
\begin{align}
\omega_r(\Phi_\mathrm{m}) &= \frac{1}{\sqrt{L_\mathrm{tot}(\Phi_\mathrm{m})C_\mathrm{tot}}}, \\
Z_r(\Phi_\mathrm{m}) &= \sqrt{\frac{L_\mathrm{tot}(\Phi_\mathrm{m})}{C_\mathrm{tot}}}.
\end{align}
Here,
\begin{equation}
L_\mathrm{tot}(\Phi_\mathrm{m}) =
\eta_L\left[L_{\mathrm{r,Sq}}(\Phi_\mathrm{m})+L_{\mathrm{r,J}}\right]
+L_g+L_k
\label{eq:l_tot}
\end{equation}
\replace{is the effective mode inductance.}{is the effective inductance of the lumped-element representation of the resonator mode.} 
The factor $\eta_L = 8/\pi^2$ accounts for mapping the distributed $\lambda/4$ resonator onto an effective lumped-element description of its fundamental mode.
The total effective capacitance is
\begin{equation}
C_\mathrm{tot}=C_{\mathrm{eq}}+C_c+C_{\mathrm{PL}}+C_{\mathrm{gates}},
\end{equation}
where $C_{\mathrm{eq}}\simeq C_{\mathrm{gnd}}/2$ is the effective capacitance to ground of the fundamental mode. From finite-element modeling and the measured zero-flux resonance frequency, we estimate $C_\mathrm{tot}\simeq20~\mathrm{fF}$. The parameters used in the effective circuit model are summarized in Table~\ref{tab:nonlin_app}.

\begin{table}[ht]
\caption{SQUID-array resonator parameters used in the effective circuit model.}
\centering
\begin{tabular}{l c c}
\hline\hline
Parameter & Symbol & Value \\
\hline
Number of unit cells & $N$ & 34 \\
Array length & $l$ & $200~\mu\mathrm{m}$ \\
Zero-flux resonator frequency & $\omega_r(0)/2\pi$ & $6.2~\mathrm{GHz}$ \\
Characteristic impedance & $Z_r(0)$ & $1.1~\mathrm{k}\Omega$ \\
SQUID junction inductance & $L_{\mathrm{Sq}}^0$ & $1.25~\mathrm{nH}$ \\
SQUID junction capacitance & $C_{\mathrm{Sq}}$ & $80~\mathrm{fF}$ \\
Large junction inductance & $L_J^\star$ & $0.11~\mathrm{nH}$ \\
Large junction capacitance & $C_J^\star$ & $800~\mathrm{fF}$ \\
Total SQUID inductance & $L_{\mathrm{r,Sq}}^0$ & $20~\mathrm{nH}$ \\
Total series-junction inductance & $L_{\mathrm{r,J}}$ & $2~\mathrm{nH}$ \\
Geometric inductance & $L_g$ & $0.3~\mathrm{nH}$ \\
Kinetic inductance & $L_k$ & $1.5~\mathrm{nH}$ \\
Effective total inductance & $L_\mathrm{tot}$ & $30.6~\mathrm{nH}$ \\
Capacitance to ground & $C_\mathrm{gnd}$ & $18.9~\mathrm{fF}$ \\
Coupling capacitance to feedline & $C_c$ & $5~\mathrm{fF}$ \\
Capacitance to DQD plunger gate & $C_{\mathrm{PL}}$ & $2~\mathrm{fF}$ \\
Capacitance to surrounding gates & $C_{\mathrm{gates}}$ & $3~\mathrm{fF}$ \\
\hline\hline
\end{tabular}
\label{tab:nonlin_app}
\end{table}

\section{Flux dependence of the SQUID-array resonance frequency and external coupling}
\label{app:flux_beta_kext}

In this section we examine the flux dependence of the fundamental resonance frequency of the  SQUID-array and the scaling of the external coupling rate $\kappa_\mathrm{ext}$.

\subsection{Effective inductance of a SQUID unit cell}

For a symmetric SQUID with negligible loop inductance, the effective critical current is
\begin{equation}
I_c(\Phi_\mathrm{m}) =
2I_{c0}\left|\cos\left(\pi\Phi_\mathrm{m}/\Phi_0\right)\right|,
\end{equation}
where $I_{c0}$ is the critical current of each junction, $\Phi_\mathrm{m}$ is the magnetic flux threading the SQUID loop, and $\Phi_0=h/2e$ is the superconducting flux quantum. The corresponding small-signal Josephson inductance is
\begin{equation}
L_\mathrm{Sq}(\Phi_\mathrm{m}) =
\frac{\Phi_0}{2\pi I_c(\Phi_\mathrm{m})}
=
\frac{L_\mathrm{Sq}^0}
{\left|\cos\left(\pi\Phi_\mathrm{m}/\Phi_0\right)\right|},
\label{eq:Lsq_flux}
\end{equation}
with $L_\mathrm{Sq}^0=L_\mathrm{Sq}(\Phi_\mathrm{m}=0)$. 
In the array, additional flux-independent inductive contributions arise from the large-area series junctions, geometric inductance, and kinetic inductance of the superconducting wiring. 
These contributions are included in the effective inductance $L_\mathrm{tot}(\Phi_\mathrm{m})$ defined in Eq.~\eqref{eq:l_tot}.

\subsection{Flux dependence of the external coupling rate}

The resonator is capacitively coupled to a microwave feedline through the coupling capacitance $C_c$, as shown in Fig.~\ref{fig:SampleAndCircuit}(b). In the weak-coupling regime, the external loss rate can be expressed as
\begin{equation}
\kappa_\mathrm{ext}=\frac{\omega_r}{Q_\mathrm{ext}}.
\end{equation}
For a capacitively coupled resonator mode, input--output theory 
gives the scaling
\begin{equation}
\kappa_\mathrm{ext}
\propto
C_c^2\,\omega_r^3\,Z_r(\omega_r)\,Z_\mathrm{TL},
\label{eq:kext_scaling}
\end{equation}
where $Z_r(\omega_r)$ is the effective impedance of the resonator and $Z_\mathrm{TL}$ is the effective transmission-line impedance of the microwave feedline coupled to the resonator through $C_c$. The proportionality constant depends on the precise mode distribution of the $\lambda/4$ resonator and contains geometry factors of order unity~\cite{Goeppl2009,Wong2017}.

From the lumped-element circuit model
\begin{equation}
Z_r(\Phi_\mathrm{m})=
\sqrt{\frac{L_\mathrm{tot}(\Phi_\mathrm{m})}{C_\mathrm{tot}}},
\qquad
\omega_r(\Phi_\mathrm{m})=
\frac{1}{\sqrt{L_\mathrm{tot}(\Phi_\mathrm{m})C_\mathrm{tot}}},
\end{equation}
one obtains
\begin{equation}
\omega_r^3(\Phi_\mathrm{m})Z_r(\Phi_\mathrm{m})
=
\frac{1}{L_\mathrm{tot}(\Phi_\mathrm{m})C_\mathrm{tot}^2}.
\end{equation}
Thus, for approximately flux-independent $C_\mathrm{tot}$ and $Z_\mathrm{TL}$, the external coupling is expected to scale as
\begin{equation}
\kappa_\mathrm{ext}(\Phi_\mathrm{m})
\propto
\frac{C_c^2 Z_\mathrm{TL}}
{L_\mathrm{tot}(\Phi_\mathrm{m})C_\mathrm{tot}^2}.
\end{equation}
Increasing the effective resonator inductance therefore lowers $\omega_r$ and reduces $\kappa_\mathrm{ext}$.

\begin{figure}[!t]
\includegraphics[height=6cm,width=\columnwidth, keepaspectratio]{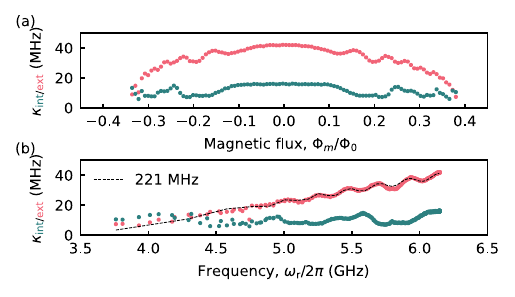}

\caption{
External and internal loss rates, $\kappa_{\mathrm{ext}}$ (red) and $\kappa_{\mathrm{int}}$ (green) as a function of (a) applied magnetic flux $\Phi_\mathrm{m}/\Phi_0$ and (b) resonator frequency $\omega_\mathrm{r}/2\pi$, extracted from the circle fit of measured resonator spectrum (see Fig.~\ref{fig:SampleAndCircuit}(e)).
The oscillations in $\kappa_{\mathrm{ext}}$ reveal a periodic modulation with a characteristic frequency of approximately $221~\mathrm{MHz}$. The black dashed line in (b) represents a sinusoidal fit with a linear background.
}

\label{fig:figure2_supp}
\end{figure}

As shown in Fig.~\ref{fig:figure2_supp}(a), which reproduces Fig.~\ref{fig:SampleAndCircuit}(f), $\kappa_\mathrm{int}$ depends only weakly on $\Phi_\mathrm{m}$, whereas $\kappa_\mathrm{ext}$ exhibits pronounced oscillations as a function of $\omega_r$, with a periodicity of approximately $221~\mathrm{MHz}$. We attribute this additional modulation to standing waves in the microwave feedline capacitively coupled to the SQUID-array resonator. These standing waves make the effective transmission-line impedance $Z_\mathrm{TL}$ frequency dependent, resulting in an oscillatory modulation of $\kappa_\mathrm{ext}$ on top of the flux dependence expected from Eq.~\eqref{eq:kext_scaling}.

\section{Frequency dependence of the light--matter coupling strength}
\label{app:g0_scaling}

In this section we derive the expected dependence of the charge–photon coupling strength $g_0$ on the fundamental resonance frequency of the SQUID-array and clarify under which assumptions a scaling with $\sqrt{\omega_r}$ is expected.

\subsection{General expression for the coupling strength}

A double quantum dot (DQD) charge qubit couples to a resonator through voltage fluctuations at the coupling node. The interaction Hamiltonian can be written as
\begin{equation}
H_\mathrm{int} = 2e\,\hat n\, \alpha\, \hat V ,
\end{equation}
where $\hat n$ is the charge operator of the DQD, $\alpha$ is the lever arm converting voltage fluctuations at the resonator node into detuning energy of the DQD, and $\hat V = V_\mathrm{zpf} (\hat a + \hat a^\dagger)$ is the quantized voltage operator of the resonator mode at the coupling point.

Projecting onto the qubit eigenbasis yields the transverse interaction
\begin{equation}
H_\mathrm{int} = \hbar g_0 (\hat a + \hat a^\dagger)\hat\sigma_x,
\end{equation}
with
\begin{equation}
g_0 = \frac{2e}{\hbar}\,\alpha\,V_\mathrm{zpf}\,\sin\theta ,
\end{equation}
where $\tan\theta = 2t/\epsilon$. At the sweet spot ($\epsilon = 0$), $\sin\theta = 1$, and the coupling reduces to
\begin{equation}
\label{eq:g0_general}
g_0(\epsilon=0) = \frac{2e}{\hbar}\,\alpha\,V_\mathrm{zpf}.
\end{equation}

The frequency dependence of $g_0$ therefore originates from (i) the zero-point voltage fluctuations $V_\mathrm{zpf}$ of the resonator mode and (ii) any implicit frequency dependence of the lever arm $\alpha$.

\subsection{Zero-point voltage fluctuations}

For a single resonator mode with effective capacitance $C_\mathrm{tot}$, the zero-point voltage is determined from the electric energy of the mode,
\begin{equation}
\frac{1}{2} C_\mathrm{tot} V_\mathrm{zpf}^2 = \frac{\hbar \omega_r}{4},
\end{equation}
which yields
\begin{equation}
\label{eq:Vzpf_Ceff}
V_\mathrm{zpf} = \sqrt{\frac{\hbar \omega_r}{2 C_\mathrm{tot}}}.
\end{equation}

Using the relations
\begin{equation}
Z_r = \sqrt{\frac{L_\mathrm{tot}}{C_\mathrm{tot}}}, \qquad
\omega_r = \frac{1}{\sqrt{L_\mathrm{tot} C_\mathrm{tot}}},
\end{equation}
one can eliminate $C_\mathrm{tot}$ and express the zero-point voltage in terms of the resonator impedance,
\begin{equation}
C_\mathrm{tot} = \frac{1}{\omega_r Z_r},
\end{equation}
leading to
\begin{equation}
\label{eq:Vzpf_Z}
V_\mathrm{zpf} = \omega_r \sqrt{\frac{\hbar Z_r}{2}}.
\end{equation}

Combining Eqs.~(\ref{eq:g0_general}) and (\ref{eq:Vzpf_Z}), the coupling strength at the sweet spot can be written as
\begin{equation}
\label{eq:g0_scaling_general}
g_0(\epsilon=0) =
\frac{2e}{\hbar}\,\alpha(\omega_r)\,
\omega_r \sqrt{\frac{\hbar Z_r(\omega_r)}{2}}.
\end{equation}

Equation~(\ref{eq:g0_scaling_general}) shows that the frequency dependence of $g_0$ is governed by both the resonator impedance $Z_r(\omega_r)$ and any possible variation of the effective lever arm $\alpha(\omega_r)$.

\subsection{Expected scaling of $g_0$ for a resonator tuned by magnetic flux}
In a SQUID-array resonator tuned via magnetic flux, the dominant effect is typically a variation of the effective inductance $L_\mathrm{tot}$, while the capacitance $C_\mathrm{tot}$ remains approximately constant. In this case,
\begin{equation}
\omega_r = \frac{1}{\sqrt{L_\mathrm{tot} C_\mathrm{tot}}}, 
\qquad
Z_r = \sqrt{\frac{L_\mathrm{tot}}{C_\mathrm{tot}}}.
\end{equation}

If $C_\mathrm{tot}$ is constant, then $L_\mathrm{tot} \propto 1/\omega_r^2$, and therefore
\begin{equation}
Z_r \propto \frac{1}{\omega_r}.
\end{equation}
Substituting into Eq.~(\ref{eq:g0_scaling_general}) yields
\begin{equation}
g_0 \propto \omega_r \sqrt{\frac{1}{\omega_r}}
= \sqrt{\omega_r},
\end{equation}
provided that $\alpha$ is frequency independent. 

Thus, a $\sqrt{\omega_r}$ dependence is expected when (i) the resonator is tuned primarily via its inductance, (ii) the effective capacitance is constant, and (iii) the coupling lever arm does not vary with frequency.

\subsection{Deviations from the expected $\sqrt{\omega_r}$ scaling}

Equation~(\ref{eq:g0_scaling_general}) also clarifies that deviations from $\sqrt{\omega_r}$ scaling naturally arise if either $Z_r$ or $\alpha$ has an additional frequency dependence.

For example:

\begin{itemize}
\item If the resonator impedance remains approximately constant during tuning ($Z_r \approx \mathrm{const}$), then
\begin{equation}
g_0 \propto \omega_r,
\end{equation}
leading to a linear dependence.

\item If capacitive loading or distributed effects modify $C_\mathrm{tot}$ as a function of frequency, then $Z_r(\omega_r)$ may deviate from the simple $1/\omega_r$ dependence expected for an ideal resonator tuned by magnetic flux.

\item The effective lever arm $\alpha$ depends on the capacitive division between the resonator node and the DQD confinement gates. 
Frequency-dependent voltage division, \add{potential frequency-dependent impedance of the DQD confinement gates,} mode-shape variations in a distributed $\lambda/4$ structure, or changes in participation ratios can introduce an additional $\omega_r$ dependence.
\end{itemize}

In general, the experimentally relevant expression is therefore Eq.~(\ref{eq:g0_scaling_general}), which shows that the scaling of $g_0$ with $\omega_r$ reflects the combined behavior of the resonator impedance and the effective voltage coupling to the DQD. Deviations from the ideal $\sqrt{\omega_r}$ dependence do not necessarily imply a modification of the intrinsic DQD dipole moment, but may instead arise from the detailed electromagnetic properties of the flux-tunable resonator and its electromagnetic environment.

\subsection{Coupling strength extracted at finite detuning}
\label{app:g_eff}
In a double quantum dot (DQD) charge qubit the light--matter interaction depends on the mixing between localized charge states. 
In the localized basis $\{|L\rangle,|R\rangle\}$ the qubit Hamiltonian reads
\begin{equation}
H_q=\frac{\epsilon}{2}\sigma_z+t\,\sigma_x ,
\end{equation}
giving the energy splitting
\begin{equation}
\hbar\omega_q(\epsilon)=\sqrt{\epsilon^2+4t^2}.
\end{equation}

The resonator electric field couples to the charge dipole and therefore modulates the detuning $\epsilon$. The interaction Hamiltonian can thus be written in the charge basis as
\begin{equation}
H_{\rm int}=\hbar g_0(\omega_r)(a+a^\dagger)\sigma_z ,
\end{equation}
where $g_0(\omega_r)$ denotes the coupling strength at the charge degeneracy point ($\epsilon=0$). 
Transforming to the qubit eigenbasis introduces the mixing angle $\theta$ defined by
\begin{equation}
\sin\theta=\frac{2t}{\hbar\omega_q}, \qquad
\cos\theta=\frac{\epsilon}{\hbar\omega_q}.
\end{equation}

The interaction Hamiltonian becomes
\begin{equation}
H_{\rm int}=\hbar g_0(\omega_r)(a+a^\dagger)\left(\cos\theta\,\sigma_z+\sin\theta\,\sigma_x\right).
\end{equation}

Only the transverse term proportional to $\sigma_x$ produces the vacuum Rabi splitting. 
Therefore the effective coupling that determines the size of the avoided crossing is
\begin{equation}
g_{\rm eff}(\epsilon,\omega_r)=g_0(\omega_r)\sin\theta
=g_0(\omega_r)\frac{2t}{\hbar\omega_q(\epsilon)}.
\label{eq:geff_general}
\end{equation}

At the charge degeneracy point $\epsilon=0$ one has $\omega_q=2t/\hbar$ and $g_{\rm eff}=g_0$.

\paragraph{Resonance condition at fixed tunnel coupling.}
In the supplementary measurement the tunnel coupling $t$ is kept constant while the resonator frequency is tuned. 
The resonance condition is reached by adjusting the detuning such that
\begin{equation}
\omega_q(\epsilon_{\rm res})=\omega_r.
\end{equation}
Using Eq.~(\ref{eq:geff_general}) this yields
\begin{equation}
g_{\rm eff,res}(\omega_r)=g_0(\omega_r)\frac{2t}{\hbar\omega_r}.
\label{eq:geff_res}
\end{equation}

Thus the measured vacuum Rabi splitting at finite detuning does not directly give the bare coupling $g_0$, but rather the reduced coupling $g_{\rm eff,res}$.

\paragraph{Frequency dependence.}
The bare coupling is determined by the zero-point voltage fluctuations of the resonator,
\begin{equation}
g_0(\omega_r)\propto \alpha(\omega_r)V_{\rm zpf}(\omega_r),
\qquad
V_{\rm zpf}=\sqrt{\frac{\hbar\omega_r}{2C_{\rm eff}}}.
\end{equation}

Combining with Eq.~(\ref{eq:geff_res}) gives
\begin{equation}
g_{\rm eff,res}(\omega_r)\propto
\alpha(\omega_r)\,V_{\rm zpf}(\omega_r)\frac{1}{\omega_r}.
\label{eq:scaling_general}
\end{equation}

For an ideal inductively tuned resonator with constant effective capacitance and lever arm one obtains
\begin{equation}
g_0(\omega_r)\propto\sqrt{\omega_r}
\quad\Rightarrow\quad
g_{\rm eff,res}(\omega_r)\propto\frac{1}{\sqrt{\omega_r}}.
\end{equation}

More generally, any frequency dependence of the resonator impedance or voltage division factor modifies this scaling. 
Equation~(\ref{eq:geff_res}) therefore provides the relation needed to extract the intrinsic coupling strength:
\begin{equation}
g_0(\omega_r)=g_{\rm eff,res}(\omega_r)\frac{\hbar\omega_r}{2t}.
\end{equation}

This correction is applied when comparing measurements performed at finite detuning with those obtained at the charge degeneracy point.
\\\\

\section{Master-equation model and input--output fits}
\label{app:master_equation}



We extract the charge--photon coupling strength and other hybrid-system parameters by fitting the measured complex normalized resonator reflection $S_{11}(\varepsilon,\omega_\mathrm{d})$ to the master-equation model in Eq.~\eqref{eq:master_equation}~\cite{ranni_decoherence_2024,dePalma2024}.



For the fit, we first determine the resonance frequency of the bare (unhybridized) resonator mode $\omega_\mathrm{r}$, along with the corresponding external and internal loss rates $\kappa_\mathrm{ext}$ and $\kappa_\mathrm{int}$. 
This is achieved by fitting a resonator spectrum at a fixed value of $\Phi_\mathrm{m}$ in a regime where the qubit frequency $\omega_\mathrm{q}$ is far detuned from $\omega_\mathrm{r}$.
For example, when analyzing the DQD detuning dependence of the hybridized mode spectrum shown in Fig.~\ref{fig:Resonator_Response}(a), we first evaluate $S_{11}$ vs. $\omega_\mathrm{d}$ at a fixed $\varepsilon/h > 2.5~\mathrm{GHz}$ and fit it to the input-output model. 
In this regime, $\omega_\mathrm{q} - \omega_\mathrm{r} \gg g_0$ holds, implying that hybridization between the resonator and the qubit is negligible. Consequently, the measured response corresponds directly to the bare resonator parameters $\omega_\mathrm{r}$ and $\kappa$.
Once the bare resonator parameters are extracted, we fit the full measured spectrum using the master equation model in Eq.~\eqref{eq:master_equation} \cite{ranni_decoherence_2024,dePalma2024}, from which we extract the intrinsic charge–photon coupling strength $g_0$ and the qubit transition frequency $\omega_\mathrm{q}$.
By performing a power sweep in the two-tone measurement configuration corresponding to zero detuning, we extract \(\Gamma_2\) by fitting the linewidth to a Lorentzian dependence, as done in \cite{Scarlino2019}, for each power point and extrapolating to zero power to obtain \(\Gamma_2\).

We define the DQD charge susceptibility as
\begin{equation}
\chi =
\frac{g_\mathrm{eff,res}}{\Delta_\mathrm{q} - i\Gamma_2},
\end{equation}
with $\Delta_\mathrm{q} = \omega_\mathrm{q} - \omega_\mathrm{d}$. We also define the resonator-drive detuning $\Delta_\mathrm{r} = \omega_\mathrm{d} - \omega_\mathrm{r}$.
Then, the reflection coefficient is expressed as
\begin{equation}
S_{11}(\Delta_\mathrm{r}) = 
\frac{
\Delta_\mathrm{r} + i (\kappa_\mathrm{ext} - \kappa_\mathrm{int})/2 +  g_\mathrm{eff,res}\chi
}
{
\Delta_\mathrm{r} + i \kappa/2 + g_\mathrm{eff,res}\chi
}
\label{eq:master_equation}
\end{equation}

\begin{figure*}[t]
\includegraphics[width=\textwidth]{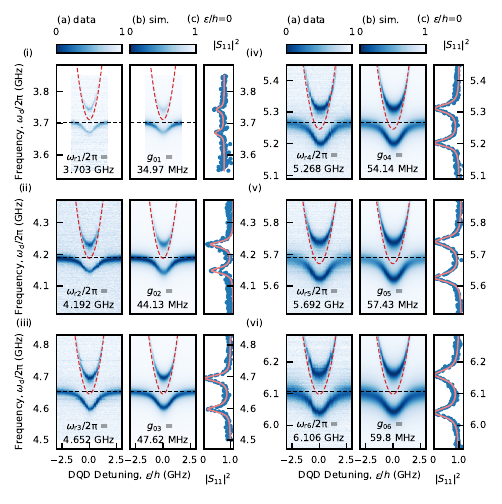}
\caption{
Vacuum Rabi mode splittings measured in reflection.
Measured (a) and simulated from the fit parameters (b) 2D reflection spectra $\mathrm{|S_{11}|^2}$ as a function of DQD detuning $\varepsilon/h$ and resonator drive frequency $\omega_d/2\pi$ for six resonator frequencies (i-vi) from which the data in Fig.~\ref{fig:Resonator_Response}(d) of the main text has been extracted, with the interdot tunnel coupling $2t$ kept close to the SQUID array resonance. (c) $|S_{11}|^2$ at $\varepsilon/h=0$ showing the vacuum Rabi splitting. The red solid line is a fit to an input-output model used to extract the coupling $g_0$.
}
\label{fig:supplementary_gvsfreq}
\end{figure*}

\section{Vacuum Rabi mode splitting at different values of inter-dot tunnel coupling}
\label{app:hybrid_spectrum_different_2t}

To systematically study the charge–photon coupling as a function of the inter-dot tunnel coupling $t$, we measure $S_{11}$ as a function of the probe frequency $\omega_\mathrm{d}$ and the DQD detuning $\varepsilon$, following the approach resulting in the data shown in Fig.~\ref{fig:Resonator_Response}(a). 
The tunnel coupling $t$ is tuned using the charge gates of the DQD.
Throughout these measurements, the charge occupation in each quantum dot is kept constant to minimize changes in the confinement potential, which could otherwise modify the charge dipole moment and consequently affect the coupling strength $g_0$ \cite{scarlino2022}.
Figure~\ref{fig:supplementary_gvsfreq}(a) shows the measured $|S_{11}|^2$ as a function of $\omega_\mathrm{d}$ and $\varepsilon$ for six different values of $t$ (see Table~\ref{tab1:para_fig3}). Although only the datasets satisfying the near-resonant condition $\omega_\mathrm{r}/2\pi \sim 2t/h$ are shown, in practice $\omega_\mathrm{r}$ is tuned using the magnetic flux $\Phi_\mathrm{m}$ into the resonance at the DQD sweet spot $\varepsilon = 0$, i.e., $\omega_\mathrm{r}/2\pi \approx 2t/h$.
Each measurement of $S_{11}(\omega_\mathrm{d}, \varepsilon)$ is fitted using the master equation model described in Eq.~\eqref{eq:master_equation} (see Appendix~\ref{app:master_equation}). The extracted parameters are then used to reproduce the corresponding $|S_{11}|^2$ spectra shown in Fig.~\ref{fig:supplementary_gvsfreq}(b) (see Table~\ref{tab1:para_fig3}). The red dashed lines indicate the fitted resonator frequency $\omega_\mathrm{r}$, while the black dashed lines correspond to the calculated qubit frequency $\omega_\mathrm{q}/2\pi = \sqrt{\varepsilon^2 + 4t^2}/h$.
The vacuum Rabi splitting at $\varepsilon = 0$ for each configuration is presented in Fig.~\ref{fig:supplementary_gvsfreq}(c), where the solid red line represents the fit to the master equation model described in Appendix~\ref{app:master_equation}.

\begin{table*}[t]
\centering
\begin{tabular}{|c|c|c|c|c|c|c|c|c|}
  \hline
   $\omega_{\rm{r}}/2\pi$[GHz] & $2t/h$ [GHz] & ${g_0}/2\pi$ [MHz] & $\kappa/2\pi$ [MHz] & $\kappa_\mathrm{ext}/2\pi$ [MHz] & $\Gamma_2/2\pi$ [MHz] & ${g_0}^2/(\kappa \Gamma_2)$ & $\delta\nu_{\rm{R}}=\kappa/2 + \Gamma_2$ [MHz] \\ \hline\hline
  6.106 & 6.098 & 59.8$\pm$0.03 & 55.22$\pm$1.03 & 39.59$\pm$0.43 & 6.83$\pm$0.56 & 9.48$\pm$0.79 & 34.44$\pm$1.07  \\ \hline
  5.692 & 5.671 & 57.43$\pm$0.02 & 42.48$\pm$0.76 & 34.57$\pm$0.36 & 6.25$\pm$0.15 & 12.42$\pm$0.37& 27.49$\pm$0.53\\ \hline
  5.268 & 5.246 & 54.14$\pm$0.03 & 36.12$\pm$0.76 & 28.04$\pm$0.35 & 4.67$\pm$0.23 & 17.39$\pm$0.92 & 22.73$\pm$0.61\\ \hline
  4.652 & 4.645 & 47.62$\pm$0.02 & 22.32$\pm$0.47 & 15.9$\pm$0.21 & 4.72$\pm$0.02 & 21.51$\pm$0.46 & 15.89$\pm$0.26\\ \hline
  4.192 & 4.188 & 44.13$\pm$0.08 & 21.86$\pm$0.9 & 11.02$\pm$0.3 & 4.33$\pm$0.1 & 20.58$\pm$0.97 & 15.26$\pm$0.55\\ \hline
  3.701 & 3.712 & 34.97$\pm$0.12 & 13.89$\pm$1.6 & 3.9$\pm$0.31 & 4.04$\pm$0.29& 21.77$\pm$2.97 & 10.99$\pm$1.09\\
  \hline
\end{tabular}
\caption{Parameters of the 6 studied points in Fig. \ref{fig:Resonator_Response}. The resonator frequency ${\mathrm{\omega_r}}$, external losses $\mathrm{\kappa_{ext}}$ and internal losses $\mathrm{\kappa_{int}}$ are extracted from fitting the resonators $S_{11}$ response when the DQD is detuned. The tunnel coupling $\mathrm{2}t/h$ and intrinsic charge-photon coupling $g_{\mathrm{0}}$ are extracted from fitting the 2D spectra to the master equation. The decoherence rate $\Gamma_2$ is estimated from the charge qubit linewidth measured using power-dependent two-tone spectroscopy. 
The cooperativity $g_\mathrm{0}^2/(\kappa_{\mathrm{tot}}\Gamma_2)$ and qubit linewidth ($\delta\nu_{\mathrm{R}}$) are calculated values.  }\label{tab1:para_fig3}

\end{table*}

\begin{figure*}[t]
\includegraphics[width=\textwidth]{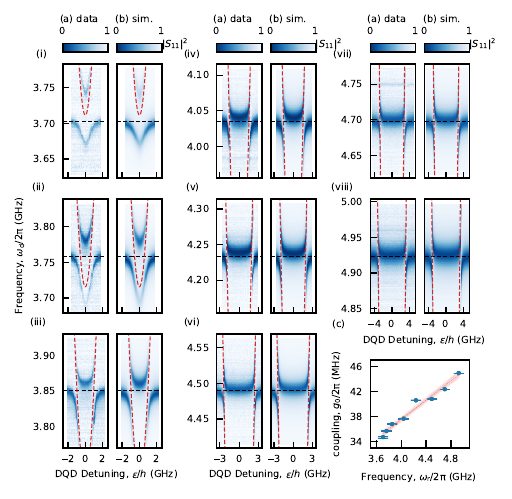}

\caption{
DQD resonator spectroscopy at fixed tunnel coupling $2t$.
Measured (a) and simulated from the fit parameters (b) $\mathrm{|S_{11}(\omega_\mathrm{d}, \varepsilon)|^2}$ as a function of DQD detuning $\varepsilon/h$ and resonator drive frequency $\omega_d/2\pi$ for eight  resonator frequencies (i-viii), with a fixed interdot tunnel coupling $2t$. 
Red dashed curves show the DQD transition frequency, while black dashed horizontal lines indicate the bare resonator frequency.
(c) Extracted DQD-resonator coupling strength $g_0/2\pi$ as a function of resonator frequency $\omega_r/2\pi$.
Blue points are values obtained from the fits, and the red line is a linear fit.
}

\label{fig:supplementary_scaling}
\end{figure*}


\section{Vacuum-Rabi mode splitting for different resonator mode frequencies at a fixed DQD configuration}
\label{app:hybrid_spectrum_different_w_r}

As shown in Fig.~\ref{fig:Resonator_Response}(d), the intrinsic charge--photon coupling strength follows a scaling $g_0 \propto \omega_\mathrm{r}$, rather than the expected $g_0 \propto \sqrt{\omega_\mathrm{r}}$ dependence.
As discussed in Appendix~\ref{app:g0_scaling}, one possible explanation for this linear scaling is a variation in the DQD dipole moment arising from different electrostatic configurations associated with varying inter-dot tunneling rates $t$.
To rule out this possibility, we fix the DQD electrostatic configuration and measure $S_{11}$ as a function of the probe frequency $\omega_\mathrm{d}$ and the DQD detuning $\varepsilon$ for different values of $\omega_\mathrm{r}$.
The corresponding $S_{11}$ measurements for eight distinct values of $\omega_\mathrm{r}$ are presented in Fig.~\ref{fig:supplementary_scaling}(a).
In these measurements, the inter-dot tunnel coupling is held constant at $2t/h \sim 3.71$~GHz.
Each measured $S_{11}$ spectrum is fitted using a master-equation model (see Appendix~\ref{app:master_equation}). The parameters extracted from these fits are then used to calculate the spectra shown in Fig.~\ref{fig:supplementary_scaling}(c) (see Table~\ref{tab1:para_fig4} for the extracted parameters).
The extracted values of $g_0$ as a function of $\omega_\mathrm{r}$ are also plotted in Fig.~\ref{fig:supplementary_scaling}(c).
Notably, although a scaling $g_0 \propto \sqrt{\omega_\mathrm{r}}$ is expected for this dataset as well (see Appendix~\ref{app:g0_scaling}), we again observe a linear dependence $g_0 \propto \omega_\mathrm{r}$.
Furthermore, the slopes $\partial g_0/\partial \omega_\mathrm{r}$ extracted from Fig.~\ref{fig:Resonator_Response}(d) and Fig.~\ref{fig:supplementary_scaling}(c) are in close agreement with each other.
This consistency indicates that the observed scaling originates from the electromagnetic properties of the distributed high-impedance resonator and its electromagnetic environment, rather than from variations in the intrinsic DQD dipole moment.


\begin{table*}[t]
\centering
\begin{tabular}{|c|c|c|c|c|c|c|c|c|c|}
  \hline
  $\omega_{\rm q}/2\pi$[GHz] &  $\epsilon/h $ [MHz] & $\omega_\mathrm{r}/2\pi$ [MHz] & $\Delta$ [MHz] & $g/2\pi$ [MHz] & $\delta\nu_\mathrm{q}(P\rightarrow0)$ [MHz] & $\Gamma_\mathrm{1}$ [MHz]  & $\Gamma_\mathrm{2}$ [MHz]  & $\Gamma_\mathrm{echo}$ [MHz] & $\Gamma_\mathrm{\phi}$ [MHz]  \\ \hline\hline
  6.030 & 0 & 5.170 & 860& 56& 7.9$\pm$0.4 & 13.3$\pm$0.7 & 12.9$\pm$2.7&-&5.4$\pm$0.9  \\ \hline
  5.020 & 0 & 5.850 & -830 & 50 & 6.2$\pm$0.3 & 6.3$\pm$0.2 & 8.3$\pm$0.9&-&4.8$\pm$0.9 \\ \hline
  4.033 & 0 & 5.059& -1026 & 37 & 4.3$\pm$0.2 & 3.4$\pm$0.04 & 6.8$\pm$0.2&3.7$\pm$0.5&5.1$\pm$0.2\\ \hline
  4.060 & 470  & 5.270 & -1210 & 37 & 9.7$\pm$0.3 & 4.0$\pm$0.1 & 11.4$\pm$1.4&3.3$\pm$0.5&-  \\ \hline
  4.114 & 810 & 5.059& -1156 & 37 & 12.2$\pm$0.4 & 3.5$\pm$0.05 & 14.8$\pm$1.8&3.9$\pm$0.3&- \\ \hline
  4.118 & 920 & 5.270& -1152& 37 & 17.5$\pm$1.4 & 3.3$\pm$0.1 & 14.7$\pm$1.2&3.7$\pm$0.5&- \\ \hline
  4.134 & 1010  & 5.268 & -1134 & 37 & 16.7$\pm$1.4 & 3.4$\pm$0.1& 16.2$\pm$2.5&4.0$\pm$0.3&- \\ \hline
  4.195 & 1155  & 5.270& -1152& 37 & 16.9$\pm$0.6 & 2.6$\pm$0.04 & 18$\pm$5.7&5.0$\pm$0.5&- \\ \hline
  3.467 & 0  & 5.267 & -1800 & - & 4.4$\pm$0.2 & 2.5$\pm$0.02 & 6.0$\pm$0.9 &-&5.1$\pm$0.9 \\ \hline
  2.987 & 0 & 5.177 & -2190 & - & - & 1.9$\pm$0.03 & 6.3$\pm$0.9&-&6.3$\pm$2.7 \\
  \hline
\end{tabular}
\caption{Parameters of the studied points in Fig. \ref{fig:Parameters}. The qubit frequency $\omega_{\mathrm{q}}$, the resonator frequency $\omega_{\mathrm{r}}$ are fixed parameters. $\Delta$ indicates the detuning between the qubit and resonator $2\pi\Delta = \omega_{\mathrm{q}} - \omega_{\mathrm{r}}$
The relaxation rate $\Gamma_1$, Ramsey rates $\Gamma_2$ and Echo rates $\Gamma_\mathrm{echo}$ are fitted while the pure dephasing is calculated according to $\Gamma_{\phi} = \Gamma_2 - \Gamma_1/2$. The linewidth $\delta \nu_{\mathrm{q}}$ is estimated from the charge qubit linewidth measured with power-dependence two-tone spectroscopy.}\label{tab1:para_fig4}
\end{table*}





\begin{figure*}[t]
\includegraphics[width=\textwidth]{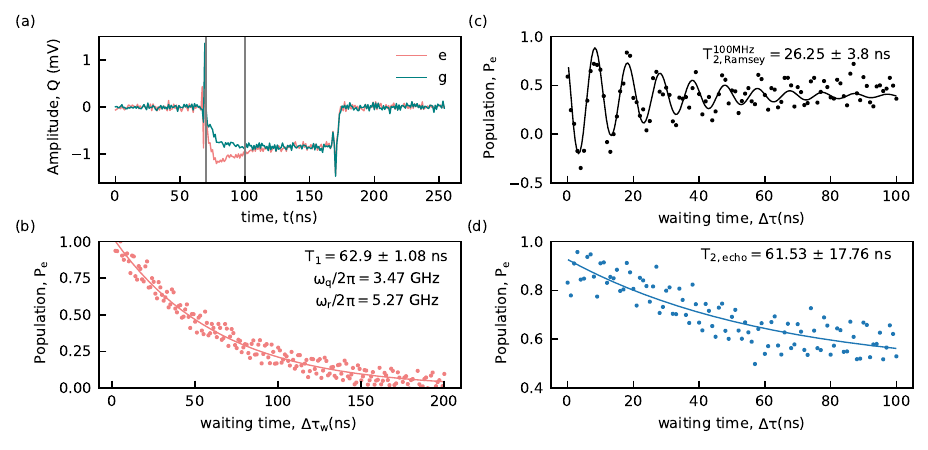}
\caption{
Time-resolved measurements of the DQD qubit at $\mathrm{\omega_q/2\pi (\epsilon/h = 0) = 2t/h } = \mathrm{3.467~GHz}$. 
(a) Time-domain response of the resonator $S_{11}$ signal with no pulse ($\mathrm{|g\rangle}$ in green) and a $\pi$ pulse applied at the qubit frequency ($\mathrm{|e\rangle}$ in red). The vertical gray lines indicate the integration window. 
(b) Measurement of the relaxation time $T_1$. The excited state population $\mathrm{P_e}$ vs the delay time $\mathrm{\Delta \tau_w}$ between the $\pi$ pulse and the readout.
(c) Ramsey fringe measurements. $\mathrm{P_e}$ vs time $\mathrm{\Delta \tau}$ separating the two $\mathrm{\pi/2}$ pulses that are applied at a $\mathrm{100~MHz}$ detuning from the qubit frequency. 
(d) Spin-echo measurement. $\mathrm{P_e}$ vs time $\mathrm{\Delta \tau}$ separating pulses. Solid lines are fits to exponential decays~\cite{Scarlino2019}
}
\label{fig:supplementary_timedomain}
\end{figure*}
\section{Time-resolved measurements}
\label{app:time_resolved_measurements}

To measure the relaxation and dephasing times, we performed a set of time-resolved measurements for each frequency configuration in Table~\ref{tab1:para_fig4}. The applied Ramsey, Rabi, and Hahn-echo pulses are explained in detail in \cite{Scarlino2019}. 
In Fig.~\ref{fig:supplementary_timedomain} a complete data set for $\mathrm{\omega_q/2\pi (\epsilon = 0) = 5.02 \, GHz}$ is shown. 
\add{
We note that the deviation of the measured Ramsey (free-induction-decay) profile (black points in Fig.~\ref{fig:supplementary_timedomain}(c)) from a purely exponential decay (black curve in the same figure) may originate from the dominance of second-order dephasing processes over first-order contributions at the sweet spot \cite{ramon2022qubit, nichol2017high}.
}


\section{Charge qubit relaxation}
\label{app:charge_qubit_relaxation_theory}

\subsection{Model}

We concentrate on a single charge in a double quantum dot, tuned close to a charge degeneracy point $(n,m) \leftrightarrow (n-1,m+1)$. The difference in onsite potential energies is characterized by the detuning $\epsilon$, while the tunneling $2t$ between the states produces an anticrossing of bonding and antibonding states. This can be modeled by the effective Hamiltonian
\[
H= \frac{\epsilon}{2}\tau_z+ t\tau_x .
\]

In the position basis of each dot, this Hamiltonian is diagonalized by $U = e^{-\frac{i}{2}\beta \tau_y}$ as
\begin{equation}
\tilde{H}=U^\dagger H U
= \frac{\hbar\Omega}{2}\tau_z ,
\end{equation}
with
\begin{equation}
\Omega= \frac{\sqrt{4t^2+\epsilon^2}}{\hbar},
\qquad
\beta=\text{arctan2}(\epsilon,2t),
\end{equation}
where $\hbar\Omega$ is the energy splitting and $\beta$ is the mixing angle between the dot states, such that the eigenstates are
\begin{equation}
\begin{split}
\ket{\psi_+} &=
\cos(\beta/2)\ket{\psi_L}
+\sin(\beta/2)\ket{\psi_R},
\\
\ket{\psi_-} &=
\sin(\beta/2)\ket{\psi_L}
-\cos(\beta/2)\ket{\psi_R}.
\end{split}
\end{equation}

Here, $\ket{\psi_{L,R}}$ are the states of the left and right dots orthogonalized using the Hund-Mulliken approximation, i.e.,
\[
\ket{\psi_L} \approx \ket{L} + \frac{s}{2}\ket{R},
\qquad
\ket{\psi_R} \approx \ket{R} - \frac{s}{2}\ket{L}.
\]

In this approximation, we used the fact that the overlap of the wavefunction is small, i.e., $s = \braket{L|R} \ll 1$, and hence corrections to the normalization can be neglected. We use the Gaussian model of the wavefunction
\begin{align}
\braket{r|R}
&\propto
\exp\!\left(-\frac{x^2+y^2}{4a^2}\right),
\\
\braket{r|L}
&\propto
\exp\!\left(
-\frac{(x+d)^2+y^2}{4a^2}
\right),
\end{align}
in which the overlap
\[
s = \exp(-d^2/8a^2)
\sim \frac{2t}{\hbar\Omega}
\]
relates the quantum dot size $a$, the distance between the dots $d$, and the ratio of the tunnel coupling $2t$ to the orbital energy $\hbar \Omega$.

The charge is coupled to the environment with the general Hamiltonian
\[
V_e=
\frac{\delta_\epsilon}{2}\tau_z
+\frac{\delta_t^r}{2}\tau_x
+\frac{\delta_t^{i}}{2}\tau_y,
\]
which gives rise to longitudinal and transverse fluctuations defined as
\begin{align}
\delta_\varphi &=
\bra{\psi_+} \hat V_e \ket{\psi_+}
-
\bra{\psi_-} \hat V_e \ket{\psi_-},
\\
\delta_R &=
\bra{\psi_+} \hat V_e \ket{\psi_-}
+
\bra{\psi_-} \hat V_e \ket{\psi_+},
\\
\delta_I &=
\bra{\psi_+} \hat V_e \ket{\psi_-}
-
\bra{\psi_-} \hat V_e \ket{\psi_+}.
\end{align}

The dephasing is related to longitudinal fluctuations $\delta_\varphi$, while relaxation is related to transverse fluctuations $\delta_R$ and $\delta_I$, which are characterized in terms of the corresponding spectral densities:
\begin{equation}
S_{xx}(\omega)
=
\int_{-\infty}^{\infty}
\text{d}\tau\,
e^{-i\omega\tau}
\langle
\delta_x^*(\tau)\delta_x(0)
\rangle,
\end{equation}
where
\[
\langle \ldots\rangle
=
\text{Tr}_E\{\rho_E \ldots \}
\]
is averaged over the bath degrees of freedom with density matrix $\rho_E$, and $x = \varphi,R,I$. The spectral density allows for computation of relaxation and dephasing times following:
\begin{align}
\frac{1}{T_1}
&=
\frac{
S_{RR}(-\Omega)+S_{RR}(\Omega)
}{4\hbar^2}
+
\frac{
S_{II}(-\Omega)+S_{II}(\Omega)
}{4\hbar^2},
\\
\chi_\varphi(t)
&=
\frac{t^2}{2\pi\hbar^2}
\int_{-\infty}^\infty
d\omega\,
S_{\varphi\varphi}(\omega)
\, \text{sinc}(\omega t/2).
\end{align}

We defined the dephasing factor as
\[
\langle \tau_x(t)\rangle
=
\tau_x(t)\exp(-\chi_\varphi[t]).
\]

In the limit of $S_{\varphi\varphi}$ dominated by low frequencies, it reduces to $\chi_\varphi = (t/T_2^*)^2$, while in the high-frequency limit it yields $\chi_\varphi = t/T_2$.

\subsection{Phonon Bath}

As one of the most relevant sources of relaxation, we consider the phonon bath, which couples to the charge through the Hamiltonian
\begin{align}
\hat V_\text{e-ph}
=
\sum_{{\bf q},p}
\Big[
\hat\rho({\bf q})
\lambda_{{\bf q},p}
\hat a_{{\bf q},p}
+
\hat\rho(-{\bf q})
\lambda_{{\bf q},p}^*
\hat a_{{\bf q},p}^\dagger
\Big].
\label{eq:hep}
\end{align}

Here, $\lambda_{{\bf q},p}$ are the coupling strengths, and
\[
\hat \rho({\bf q})
=
\int d{\bf r}\,
e^{-i{\bf q}\cdot{\bf r}}
\hat\rho ({\bf r})
\]
is the Fourier transform of the electronic density operator. The sum runs over all allowed phonon wave vectors ${\bf q}$ and includes three polarizations (one longitudinal and two transverse), labeled by $p = l, t_1, t_2$.

When substituted into the definition of the spectral density of transverse fluctuations, we find
\begin{align}
S_{R/I}(\omega)
&=
\sum_{{\bf q},{\bf q'},p,p'}
\lambda_{{\bf q}, p}
\lambda_{{\bf q'}, p'}^*
\nonumber
\\
&\quad\times
\big[
\rho_{+-}({\bf q})
\rho_{+-}(-{\bf q'})
\pm
\rho_{-+}({\bf q})
\rho_{-+}(-{\bf q'})
\big]
\nonumber
\\
&\quad\times
\big[
S_{a_{{\bf q},p}a^\dagger_{{\bf q'},p'}}(\omega)
+
S_{a^\dagger_{{\bf q},p}a_{{\bf q'},p'}}(\omega)
\big].
\end{align}

The matrix elements of the density operator are
\begin{align}
\rho_{+-}({\bf q})
&=
\rho_{+-}^*(-{\bf q})
\nonumber\\
&=
\bra{\psi_+}
\int d{\bf r}\,
e^{-i{\bf q}\cdot{\bf r}}
\hat \rho ({\bf r})
\ket{\psi_-}
\nonumber\\
&\approx
-2
e^{-\frac{a^2(q_x^2+q_y^2)}{2}}
e^{i\frac{d q_x}{2}}
\sin\beta
\nonumber\\
&\quad\times
\left[
\frac{\epsilon}{\hbar\omega}
\sin^2\!\left(
\frac{d q_x}{4}
\right)
-
i\sin\left(
\frac{d q_x}{2}
\right)
\right].
\end{align}

Thus,
\[
S_R(\omega)
\propto
2|\rho_{+-}(\mathbf{q})|^2,
\qquad
S_I(\omega)=0,
\]
since
\[
\rho_{-+}(\mathbf{q})
=
\rho_{+-}^*(\mathbf{q}).
\]

The spectral density of the phonon bath reads
\begin{align}
&S_{a_{{\bf q},p}a^\dagger_{{\bf q'},p'}}(\omega)
\nonumber\\
&=
\int_{-\infty}^\infty
d\tau\,
e^{-i(\omega-\omega_{\bf q}) \tau}
\langle
a_{{\bf q},p}
a^\dagger_{{\bf q'},p'}
\rangle
\nonumber\\
&=
2\pi
\delta_{{\bf q} {\bf q'}}
\delta_{pp'}
[1+N(\omega_{\bf q})]
\delta(\omega-\omega_{\bf q}),
\end{align}
and analogously
\begin{align}
S_{a^\dagger_{{\bf q},p}a_{{\bf q'},p'}}(\omega)
=
2\pi
\delta_{pp'}
\delta_{{\bf q} {\bf q'}}
N(\omega_{\bf q})
\delta(\omega+\omega_{\bf q}),
\end{align}
where
\[
N(x)
=
\left(
e^{{\hbar x}/{k_BT}}-1
\right)^{-1},
\]
and
\[
a_{{\bf q},p}(\tau)
=
a_{{\bf q},p}
e^{-i\omega_{\bf q} \tau}.
\]

We now compute the relaxation rate for a GaAs double quantum dot. The coupling strengths read as
\begin{equation}
\lambda_{{\bf q},p}
=
M^{(p)}_{\rm ph}
\sqrt{
\frac{\hbar}{
2\rho {\cal V} \omega_{{\bf q},p}
}
}.
\end{equation}

Here, $\rho$ is the mass density ($\rho = 5.3 \times 10^3$ kg/m$^3$ for GaAs), ${\cal V}$ is the normalization volume, and we assume coupling to acoustic phonons with an isotropic linear dispersion relation at all energies of interest, i.e.,
\[
\omega_{{\bf q},p} = v_p q,
\]
where $v_p$ is the polarization-dependent sound velocity.

The constant
\[
M^{(p)}_{\rm ph}
=
M^{(p)}_{\rm pe}
+
M_{\rm def}
\]
contains contributions from piezoelectric and deformation potential couplings~\cite{bruus1993}, with
\[
M^{(p)}_{\rm pe}
=
ieh_{14}A_{{\bf q},p},
\qquad
M_{\rm def}
=
\Xi q \,\delta_{p,l}.
\]

Here, $h_{14} = 1.38\times 10^9$ V/m and $\Xi = 13.7$~eV for GaAs~\cite{bruus1993}. The piezoelectric coupling involves anisotropy factors~\cite{bruus1993}, expressed as
\begin{align}
A_{{\bf q},l}
&=
3 \cos(\theta)\sin^2(\theta)
\sin(2\phi+2\chi),
\\
A_{{\bf q},t1}
&=
-\frac{1}{2}
[1+3\cos(2\theta)]
\sin(\theta)
\sin(2\phi+2\chi),
\\
A_{{\bf q},t2}
&=
-\sin(2\theta)\cos(2\phi+2\chi).
\end{align}

Here, $\theta = 0$ corresponds to ${\bf q}$ parallel to the $z$-axis, and $\phi$ gives the azimuthal angle of ${\bf q}$ in the $xy$-plane. The angle $\chi$ is the angle between the double-dot axis and the crystallographic (100) direction.

We replace the sum over wave vectors by an integral,
\begin{align}
\sum_{\bf q}
&\to
\int \frac{d{\bf q}}{(2\pi)^3}
\nonumber\\
&=
V
\int_0^\infty q^2 dq
\int_0^\pi \sin\theta d\theta
\int_0^{2\pi}
\frac{d\phi}{(2\pi)^3},
\end{align}
and use
\[
\hbar\Omega = \sqrt{\epsilon^2 + 4t^2}
\]
to find the relaxation rate
\begin{align}
\Gamma_{t\alpha}
&=
\frac{
\coth(\hbar\Omega/k_BT)
}{
\pi^2\hbar^2\rho v_t^3
}
\frac{2t^2}{\hbar\Omega}
I_{t\alpha}(\Omega,d,a),
\\
\Gamma_l
&=
\frac{
\coth(\hbar\Omega/k_BT)
}{
\pi^2 \hbar^2\rho v_l^3
}
\frac{2t^2}{\hbar\Omega}
I_l(\Omega,d,a).
\end{align}

We compute the integrals $I_{t\alpha}$ and $I_l$ in the low-energy limit by expanding to leading order in $E$ with energy scales smaller than
\[
\Delta_{a}^{(p)} = \hbar v_p / a \sim 100\mu\text{eV},
\qquad
\Delta_{d}^{(p)} = \hbar v_p / d \sim 10\mu\text{eV},
\]
which are typical for GaAs devices:
\begin{align}
I_{t\alpha}
&\approx
\int d\theta d\phi\,
\sin\theta
(eh_{14})^2
A_{{\bf q},t\alpha}^2
\nonumber\\
&\quad\times
\left(
\frac{
d \hbar\Omega
\sin\theta\cos\phi
}{
2\hbar v_t
}
\right)^2,
\\
I_l
&\approx
\int d\theta d\phi\,
\sin\theta
\left[
(eh_{14})^2 A_{{\bf q},l}^2
+
\frac{\Xi^2 \Omega^2}{\hbar^2 v_l^2}
\right]
\nonumber\\
&\quad\times
\left(
\frac{
d \hbar\Omega
\sin\theta\cos\phi
}{
2\hbar v_t
}
\right)^2.
\end{align}

Evaluating analytically, we find
\begin{align}
I_{t1} = I_{t2}
&\approx
\frac{32\pi}{105}
(eh_{14})^2
\left(
\frac{
d\hbar \Omega
}{
2\hbar v_t
}
\right)^2,
\\
I_l
&\approx
\pi
\left[
\frac{16}{35}(eh_{14})^2
+
\frac{4}{3}
\frac{\Xi^2\hbar^2 \Omega^2}{\hbar^2 v_l^2}
\right]
\left(
\frac{
d \hbar\Omega
}{
2\hbar v_t
}
\right)^2.
\end{align}

Substituting into the relaxation rates, and in the limit of low temperatures, we find
\begin{align}
\Gamma_{t1} + \Gamma_{t2}
&=
\frac{32}{105}
\frac{
(eh_{14})^2 t^2 d^2 \hbar\Omega
}{
\pi \hbar^4 \rho v_t^5
},
\\
\Gamma_l
&=
\frac{
2 t^2 d^2 \hbar\Omega
}{
\pi \hbar^4 \rho v_t^5
}
\left[
\frac{4(eh_{14})^2}{35}
+
\frac{\Xi^2\hbar^2 \Omega^2}
{3\hbar^2 v_l^2}
\right].
\end{align}
At zero detuning $\hbar\Omega=2t$,
\begin{align}
\Gamma_{t1} + \Gamma_{t2}
&=
\frac{64}{105}
\frac{
(eh_{14})^2 d^2
}{
\pi \hbar^4 \rho v_t^5
}
t^3,
\\
\Gamma_l
&=
\frac{
4 d^2
}{
\pi \hbar^4 \rho v_t^5
}
\left[
\frac{4(eh_{14})^2}{35}
+
\frac{4\Xi^2 t^2}
{3\hbar^2 v_l^2}
\right]
t^3,
\end{align}
resulting in the measured scaling of the relaxation rate
\[
\Gamma_1 = 1/T_1 \propto t^3.
\]

\bibliographystyle{apsrev4-1}
\bibliography{References}
\widetext
\clearpage

\end{document}